\documentclass[prc,twocolumn,superscriptaddress,showpacs,floatfix]{revtex4}

\usepackage[dvips]{graphicx}
\usepackage{mathrsfs}
\usepackage{amssymb}
\usepackage{lscape}

\newcommand{\beq}{\begin{equation}}
\newcommand{\eeq}{\end{equation}}
\newcommand{\half}{\mbox{$\frac{1}{2}$}}
\newcommand{\qua}{\mbox{$\frac{1}{4}$}}
\newcommand{\twelve}{\mbox{$\frac{1}{12}$}}
\newcommand{\six}{\mbox{$\frac{1}{6}$}}

\newcommand{\eight}{\mbox{$\frac{1}{8}$}}
\newcommand{\tfour}{\mbox{$\frac{1}{24}$}}

\newcommand{\bH}{\mbox{$\bar{H}$}}
\newcommand{\p}{\prime}

\begin{document}

\title{Coupled-cluster calculations for valence systems around $^{16}$O}

\author{J.~R.~Gour}
\affiliation{Department of Chemistry,
Michigan State University, East Lansing, MI 48824, USA}

\author{P.~Piecuch}
\affiliation{Department of Chemistry,
Michigan State University, East Lansing, MI 48824, USA}
\affiliation{Department of Physics and Astronomy,
Michigan State University, East Lansing, MI 48824, USA}

\author{M.~Hjorth-Jensen}
\affiliation{Department of Physics,
University of Oslo, N-0316 Oslo, Norway}
\affiliation{Center of Mathematics for Applications,
University of Oslo, N-0316 Oslo, Norway}
\affiliation{Department of Physics and Astronomy,
Michigan State University, East Lansing, MI 48824, USA}

\author{M.~W{\l}och}
\affiliation{Department of Chemistry,
Michigan State University, East Lansing, MI 48824, USA}

\author{D.~J.~Dean}
\affiliation{Physics Division, Oak Ridge National Laboratory,
P.O. Box 2008, Oak Ridge, TN 37831, USA}
\affiliation{Center of Mathematics for Applications,
University of Oslo, N-0316 Oslo, Norway}

\date{\today}

\begin{abstract}
We study the ground and low-lying 
excited states of $^{15}$O, $^{17}$O, $^{15}$N, and 
$^{17}$F using modern two-body nucleon-nucleon interactions and the 
suitably designed variants of the {\em ab initio} equation-of-motion
coupled-cluster theory aimed at an
accurate description of systems with valence particles and holes. 
A number of properties of $^{15}$O, $^{17}$O, $^{15}$N, and $^{17}$F,
including ways the energies of ground and excited states
of valence systems around $^{16}$O
change as functions of the number of nucleons, are correctly reproduced
by the equation-of-motion coupled-cluster calculations.
Within a harmonic oscillator basis and large effective model spaces, our results are
converged for the chosen two-body Hamiltonians. Thus, all disagreements
with experiment are, most likely, due to the degrees of freedom such 
as three-body
interactions not accounted for in our effective two-body Hamiltonians.
In particular, the calculated binding energies of $^{15}$O/$^{15}$N and $^{17}$O/$^{17}$F
enable us to rationalize the discrepancy between the experimental and recently
published [Phys. Rev. Lett. {\bf 94}, 212501 (2005)] equation-of-motion
coupled-cluster excitation energies for the $J^{\pi}=3^{-}$ state of $^{16}$O.
The results demonstrate the feasibility of the
equation-of-motion coupled-cluster methods to deal with
valence systems around closed-shell nuclei
and to provide precise results for systems beyond $A =16$.
\end{abstract}

\maketitle

\section{Introduction} \label{sec:intro}

The way shell closures and single-particle energies evolve as functions of the
number of nucleons is presently one of the greatest challenges
to our understanding of the basic features of nuclei. The properties of
single-particle energies and states with a strong quasi-particle content
along an isotopic chain are moreover expected to be strongly influenced by the  
nuclear spin-orbit force. The latter can be retraced to contributions
from both two-body and three-body models of the nuclear forces (see, for example,
Refs.~\cite{pp1993,ab1981}).
A fully microscopic {\em ab initio} description of masses, shell
closures, excited states, and single-particle
energies in terms of the underlying nuclear forces is an unresolved problem
in nuclear physics that awaits a satisfactory and computationally tractable
solution. 

For light nuclei with mass numbers $A \le 12$, both Green's function Monte Carlo
methods \cite{pieper02} and large-scale no-core shell-model calculations
\cite{navratil02} provide almost converged benchmarks for selected two-
and three-body Hamiltonians, where typically the models for the
two-body nucleon-nucleon interactions reproduce
the available scattering data, while the three-body interaction models are normally
fitted to reproduce the binding energies of selected nuclei.
The agreement with experimental data for many light nuclei is in these
calculations quite reasonable. Unfortunately, for medium-mass and
heavier nuclei the dimensionality of the corresponding many-particle problem
becomes intractable by the Green's function Monte 
Carlo methods and {\em ab initio}
no-core shell-model techniques, and one typically has to resort to a simplified
shell-model description within a smaller space,
the so-called model space. In order to solve the corresponding many-body
Schr\"odinger equation, one needs 
then to derive effective two and/or three-body 
interactions for the chosen small model space. Many-body perturbation
theory is normally employed to derive effective interactions \cite{hko95}, but 
unless these interactions are fitted to reproduce selected
properties of nuclei \cite{brown2001,otsuka2004}, 
one cannot correctly recover the experimentally derived single-particle and excitation
energies and shell closures (see, for example, Ref.~\cite{brown2002}).

Two key points make it imperative to investigate new theoretical 
methods that will allow for an accurate description of 
closed- as well as open-shell nuclei with $A \gg 12$. First,
present and proposed nuclear structure research facilities
will open significant territory into regions of medium-mass and heavier nuclei,
where the majority of the studied nuclei will be open-shell systems and where 
many of the nuclei produced in experiment will be unstable or short-lived.
Second, existing shell-model and Green's function Monte Carlo techniques 
have prohibitive computer costs that scale factorially or exponentially 
with the system size.  
In addition to an increased dimensionality,
one needs to account for the fact that many of the medium-mass and heavier
nuclei can be weakly
bound and couple to resonant states. Moreover, in order to
examine new nuclei that have not been 
discovered or studied before, one may not be able to
rely on fitting the effective Hamiltonians to the experimental data for the
known nuclei, as has been traditionally done for many years.
Microscopic {\em ab initio} methods, in which
nuclear properties are obtained from the underlying nucleon-nucleon 
interactions,
will become increasingly important as the new information
about the medium-mass and heavier nuclei is obtained in various experiments.
In addition to these practical 
aspects, {\em ab initio} calculations
of nuclear properties, including, for 
example, the way the binding and excitation
energies change as a function of the number of nucleons around closed-shell
nuclei, may provide important new insights into our understanding of nuclear
forces.

Clearly, if we wish to extend {\em ab initio} methods
to nuclei with $A \gg 12$, we have to consider alternatives to
the existing Green's function Monte Carlo and no-core shell-model techniques.
In this work, we focus on coupled-cluster theory
\cite{coester,coesterkummel,cizek,palduscizek}, which is a promising
candidate for the development of practical methods for
fully microscopic {\em ab initio} studies of nuclei in the $A \gg 12$
mass region. As has been demonstrated over and over again in numerous quantum chemistry
applications (see, e.g., Refs.~\cite{chem_rev1,chem_rev2,chem_rev3,chem_rev4,chem_rev5,chem_rev6,%
chem_rev7,piecuch2002,%
piecuch2003,piecuch2004,piecuch2005} for selected reviews),
coupled-cluster methods are capable of providing a precise description
of many-particle correlation effects at the relatively low computer costs,
when compared to shell-model or configuration interaction techniques aimed at
similar accuracies. Based on the remarkable success of coupled-cluster
methods in chemistry and molecular physics, where one has to
obtain a highly accurate description of many-electron correlation effects,
we believe that the field of nuclear physics may significantly benefit
from the vast experience in the development of accurate and
computationally efficient coupled-cluster approximations and algorithms
by quantum chemists.

Although historically coupled-cluster theory
originated in nuclear physics \cite{coester,coesterkummel},
its applications to the nuclear many-body problem have been relatively rare
(see, e.g., Refs.~\cite{pr1978,kummel2002,aggravatedrectifier97}, and references therein),
particularly when compared
to quantum chemistry. For many years, part of the problem has been an inadequate
understanding of nucleon-nucleon interactions and lack of adequate
computer resources in the 1970s and 1980s. This situation has changed only
in the last few years. The successful
construction of realistic nucleon-nucleon potentials
(cf., e.g., Refs.~\cite{v18,cdbonn2000,entem2002,entem2003}) and
spectacular improvements in computer
technology have led to renewed interest in
applying coupled-cluster methods in {\em ab initio} nuclear physics calculations.
In particular, using bare interactions, Mihaila and
Heisenberg performed impressive
coupled-cluster calculations for the binding
energy and the electron scattering form factor of $^{16}$O
\cite{bogdan1,bogdan2,bogdan3,bogdan4}. We have taken an alternative
route and combined a few basic coupled-cluster techniques, developed earlier in the context
of electronic structure studies by quantum chemists,
with the renormalized form of the Hamiltonian
to determine ground and selected excited states of $^4$He and $^{16}$O
\cite{dean04,kowalski04,wloch05,dean05,epja2005,jpg2005},
demonstrating promising results when compared with the results
of the exact shell-model diagonalization in the same model space \cite{kowalski04}
and, at least for some properties, with the experimental data
\cite{dean04,wloch05,dean05,epja2005,jpg2005}.
In particular, in our most recent study of the ground and excited states
of $^{16}$O \cite{wloch05},
we obtained fully converged results which are
very close to those obtained with more
expensive large-scale no-core shell-model calculations of
Navr{\'a}til and collaborators \cite{navratil2004}. This has been possible
thanks to the use of the elegant diagram factorization techniques
developed by quantum chemists \cite{sak1}, which
lead to almost perfectly vectorized \cite{piecuch2005,ccgamess,wloch2005}
and highly scalable parallel \cite{dean04} computer codes, enabling routine
calculations for systems in the $A \sim 20$ region with
large single-particle basis sets,
including seven or even eight major oscillator shells (336 and 480 single-particle states,
respectively).
It should be emphasized that although we are still in the early
stages of developing a library of efficient, general-purpose
coupled-cluster programs
for nuclear structure applications, there are already several differences between
our approach to nuclear coupled-cluster calculations and the
approach pursued by Mihaila and Heisenberg \cite{bogdan1,bogdan2,bogdan3,bogdan4}.
First of all, Mihaila and Heisenberg used bare interactions,
making the convergence with the number of single-particle basis
states very slow, whereas we use the renormalized form of the Hamiltonian
exploiting, for example, a no-core G-matrix theory \cite{dean04}, which leads to
a rapid convergence of binding and excitation energies and other nuclear
properties with the number of major oscillator shells in a basis set
\cite{dean04,wloch05,dean05,epja2005,jpg2005}.
Second, we are able to calculate ground as well as excited states, not just the
ground-state properties of closed-shell nuclei examined by Mihaila and Heisenberg.
Finally, as mentioned above, our
coupled-cluster computer codes have been developed using diagram factorization techniques,
which minimize the CPU operation count, rather than the
commutator expansions used by Mihaila and Heisenberg and the Bochum school.
The coding style adopted by us is similar to that practiced by
the lead developers of coupled-cluster methods in chemistry.
In particular, we put an emphasis on the general-purpose
character of our codes, meaning that
the only essential input variables are the number of particles and the
matrix elements of the Hamiltonian in some single-particle basis set.

Our initial coupled-cluster calculations \cite{dean04,kowalski04,wloch05,dean05,epja2005,jpg2005}
have focused on closed-shell nuclei. However, the long-term objective
of our research program is to study open-shell nuclei with one or more valence nucleons.
We would like, for example, to examine how the binding and excitation
energies vary with the number of nucleons in valence systems around closed-shell
nuclei. This is particularly interesting, when we examine the $A=15$ and $A=17$
nuclei around $^{16}$O. For example,
the splittings between the $(3/2)_1^-$ and $(1/2)_1^-$ states
in $^{15}$O and $^{15}$N and the splittings between the
$(3/2)^+_1$ and $(5/2)^+_1$ states in $^{17}$O and $^{17}$F
should arise from the nuclear spin-orbit force, which may or may not be affected
by three-nucleon interactions. One would like to examine such issues
by comparing the results of converged {\em ab initio} calculations employing
two-body interactions with the experimental energy spacings. This
requires, however, an appropriate extension of the
usual single-reference
ground-state coupled-cluster theory \cite{coester,coesterkummel,cizek,palduscizek}
to ground and excited states of valence systems around closed-shell nuclei.

In this paper, we examine, for the first time,
the applicability of two quantum-chemistry-inspired
coupled-cluster approaches, referred to as the particle-attached (PA)
(in chemistry, electron-attached
or EA \cite{oureom,eaccsd1,eaccsd2,eaccsdt1,jeffeaip}) and
particle-removed (PR)
(in chemistry, ionized or IP \cite{chem_rev3,oureom,jeffeaip,%
ipccsd2,ipccsd3,ipccsd4,ipccsdt1,ipccsdt2,ipccsdt-3})
equation-of-motion coupled-cluster (EOMCC) methods \cite{oureom,eomcc1,eomcc3},
in the converged calculations of the binding and excitation energies
of the $A=15$ ($^{15}$O, $^{15}$N) and $A=17$ ($^{17}$O, $^{17}$F) nuclei.
For these calculations, we use modern
nucleon-nucleon interactions derived from the
effective-field theory \cite{eft,bira1999},
such as N$^3$LO \cite{entem2003}, and their slightly older
phenomenological counterparts, including the
charge-dependent Bonn interaction model (CD-Bonn)
\cite{cdbonn2000} and
the $V_{18}$ model of the Argonne group \cite{v18}.
In the PA- and PR-EOMCC methods, one calculates ground and excited states
of the $(A+1)$- and $(A-1)$-particle systems by
diagonalizing the similarity-transformed Hamiltonian of the
coupled-cluster theory, resulting from the ground-state calculations
for the $A$-particle closed-shell nucleus in the
relevant $(A+1)$- and $(A-1)$-particle subspaces
of the Fock space. As shown in this paper, the PA- and PR-EOMCC
approaches provide us with practical computational techniques
for potentially accurate {\em ab initio} studies of valence systems around
the closed-shell nuclei that may provide several important insights
into the effects of the underlying nucleon-nucleon interactions
on the calculated properties of such systems. In addition to
the converged PA- and PR-EOMCC results for the $A=15$ and $A=17$ nuclei obtained with
three different types of nucleon-nucleon interactions (N$^3$LO, CD-Bonn, and $V_{18}$),
we provide several details of the PA-EOMCC, PR-EOMCC, and underlying
ground-state coupled-cluster calculations, including the factorized forms
of the relevant amplitude equations that lead to highly efficient, fully vectorized
computer codes applicable to large single-particle basis sets and masses
in the $A \sim 20 - 40$ region.

This paper is divided into four sections. In Sec.~\ref{sec:formalism},
we present our formalism
for deriving an effective two-body Hamiltonian for coupled-cluster
calculations, which takes into account 
short-range nucleon-nucleon correlations, and present the details
of the PA-EOMCC and PR-EOMCC theories that enable us to deal with
valence systems around closed-shell nuclei within the framework of the single-reference
coupled-cluster formalism. The results of PA-EOMCC and  PR-EOMCC calculations for
the $^{15}$O, $^{17}$O, $^{15}$N, and
$^{17}$F nuclei are discussed in Sec.~\ref{sec:results} and the conclusions
and perspectives are outlined in Sec.~\ref{sec:conclusions}.
The factorized forms of the PA-EOMCC and PR-EOMCC equations for
the $(A \pm 1)$-particle systems and the corresponding ground-state
coupled-cluster equations, exploited in this work, are given in the Appendix.

\section{Coupled-Cluster Equations for Valence Systems}\label{sec:formalism}

This section serves the aim of presenting the coupled-cluster theories
for open-shell nuclei used in this work, with an emphasis on the
PA-EOMCC and PR-EOMCC methods mentioned in the Introduction.
These methods are designed to handle systems with one valence
particle or one valence hole. Since our effective model spaces for
the coupled-cluster calculations, which enable us to obtain converged
PA-EOMCC and PR-EOMCC results, involve up to eight major
harmonic oscillator shells (480 uncoupled single-particle basis states)
and 15 to 17 explicitly correlated nucleons, we focus on the computationally efficient
formulation of the PA-EOMCC and PR-EOMCC methods that makes such large-scale
nuclear structure calculations manageable.

Our theoretical considerations start with the introduction of an appropriate
two-body effective interaction for the large-scale coupled-cluster calculations.
This is because the majority of modern nucleon-nucleon interactions, including
the N$^3$LO, CD-Bonn, and $V_{18}$ potentials examined in this study, include
repulsive cores that would require calculations in extremely large model
spaces to reach converged results \cite{bogdan1,bogdan2,bogdan3,bogdan4}.
In order to remove the hard-core part of the interaction from the problem and allow
for realistic calculations in manageable model spaces with seven or eight
major oscillator shells, one has to renormalize the bare interactions.
The relevant information about the method used by us to renormalize
the N$^3$LO, CD-Bonn, and $V_{18}$ Hamiltonians and to generate
the final effective Hamiltonians corrected for the center-of-mass contaminations,
which can be used in the PA-EOMCC, PR-EOMCC, and other coupled-cluster calculations,
are discussed in Sec.~\ref{subsec:gmatrix}.

As mentioned in the Introduction, the PA- and PR-EOMCC methods for
valence systems are based on an idea of
diagonalizing the similarity-transformed Hamiltonian of the
coupled-cluster theory, resulting from the ground-state calculations
for the $A$-particle closed-shell nucleus, in the
relevant $(A+1)$- and $(A-1)$-particle subspaces
of the Fock space. Thus, before introducing the details of the
PA- and PR-EOMCC calculations for the $(A+1)$- and $(A-1)$-particle systems,
we provide the most essential information about the underlying closed-shell
coupled-cluster calculations that precede the PA- and PR-EOMCC steps.
This is done in Sec.~\ref{subsec:singlerefccsd}.

Finally, in Sec.~\ref{subsec:eomccsd}, we discuss the most essential details
of the PA- and PR-EOMCC calculations for the ground- and excited states
of the $(A+1)$- and $(A-1)$-particle valence systems
around the closed-shell $A$-particle nucleus. The final working equations for
the cluster and excitation amplitudes, which define the
coupled-cluster, PA-EOMCC, and PR-EOMCC approximations implemented
in this work and which lead to highly efficient computer codes, are
shown in the Appendix.

\subsection{Effective Two-Body Interaction for Coupled-Cluster
Calculations}\label{subsec:gmatrix}

In the PR-EOMCC calculations for $^{15}$O and $^{15}$N, the PA-EOMCC
calculations for $^{17}$O and $^{17}$F, and the underlying
closed-shell coupled-cluster calculations for the ground-state
of $^{16}$O which precede the PR-EOMCC and PA-EOMCC calculations,
we used the following three nucleon-nucleon interactions:
N$^3$LO \cite{entem2003}, CD-Bonn \cite{cdbonn2000}, and $V_{18}$ \cite{v18}.
The Coulomb interaction was included in all of our calculations
(to distinguish between $^{15}$O/$^{17}$O and
$^{15}$N/$^{17}$F). In order to remove the hard-core
part of the interaction, which would require calculations in extremely large model
spaces consisting of dozens of major oscillator shells to
reach reasonably converged results \cite{bogdan1,bogdan2,bogdan3,bogdan4}, and enable
realistic calculations in manageable model spaces, we follow the
procedure exploited in our earlier work \cite{dean04,kowalski04,wloch05,dean05,epja2005,jpg2005}.
Thus, we renormalize
the Hamiltonian through a no-core $G$-matrix procedure, described in considerable detail in
Refs. \cite{hko95,dean04}. The no-core $G$-matrix approach introduces a
starting-energy ($\omega$) dependence in the effective two-body matrix
elements $G(\omega)$
defining the renormalized two-body interactions
(obtained by analyzing the
exactly solvable proton-proton, proton-neutron, and neutron-neutron two-body problems),
but much of the $\omega$ dependence can be eliminated through the
use of the Bethe-Brandow-Petschek theorem
\cite{bbp63} and the appropriate summation of the class of
folded diagrams to infinite order at a given starting
energy (see Refs.~\cite{hko95,dean04} for further information).
For nuclei like $^{16}$O, the dependence on the chosen starting energy
$\omega$ is weak (almost none when seven or eight major
oscillator shells are employed \cite{dean04}).
It introduces an uncertainty of $0.1-0.2$ MeV per particle for the binding energies.

After renormalizing bare interactions with the $G$-matrix approach,
our effective Hamiltonian is given by the formula
\beq
H_{\rm eff}(\omega)=H_{0}+G(\omega),
\label{hprime}
\eeq
where
$H_{0}$ is the total kinetic energy of the nucleons. To complete the process of preparing the
Hamiltonian for coupled-cluster calculations, we correct the renormalized
Hamiltonian $H_{\rm eff}(\omega)$, Eq. (\ref{hprime}),
resulting from exploiting the no-core $G$-matrix procedure,
for center-of-mass
contaminations using the expression
\begin{eqnarray}
H \equiv H(\omega,\beta_{\rm CoM})
& = & H_{\rm eff}(\omega)+\beta_{\rm CoM} H_{\rm CoM}
\nonumber
\\
& = & z_{\alpha}^{\beta} a^{\alpha} a_{\beta}
+ \qua \, v_{\alpha\beta}^{\gamma\delta} a^{\alpha} a^{\beta} a_{\delta} a_{\gamma} ,
\label{hfinal}
\end{eqnarray}
where $z_{\alpha}^{\beta} = \langle \alpha | z | \beta \rangle$
and $v_{\alpha\beta}^{\gamma\delta} =
\langle \alpha\beta | v | \gamma\delta \rangle
- \langle \alpha\beta | v | \delta\gamma \rangle$ are
the relevant one- and two-body
matrix elements in a single-particle basis set $\{ |\alpha\rangle \}$ and
$a^{\alpha}$ ($a_{\alpha}$) are the usual creation (annihilation) operators. Here and
elsewhere in the present paper, we use the Einstein summation
convention over repeated upper and lower indices.
The parameter $\beta_{\rm CoM}$
is chosen such that the expectation value of
the center-of-mass Hamiltonian $H_{\rm CoM}$ with the ground-state
coupled-cluster wave function, $\langle H_{\rm CoM} \rangle$,
obtained for the $\beta_{\rm CoM}$-dependent
Hamiltonian $H$, Eq. (\ref{hfinal}), is 0.0 MeV. This can be
done by relying, for example, on the Hellmann-Feynman theorem and
calculating $\langle H_{\rm CoM} \rangle$ as the first derivative
of the coupled-cluster energy with respect to $\beta_{\rm CoM}$.
Once we know the values of $\langle H_{\rm CoM} \rangle$ at
various $\beta_{\rm CoM}$ values, we can easily
identify the optimum $\beta_{\rm CoM}$ value at which $\langle H_{\rm CoM} \rangle$
becomes 0.0 MeV \cite{dean99}.
As pointed out in our earlier papers \cite{wloch05,epja2005,jpg2005},
one of the advantages of this procedure is the ease
of separation of intrinsic and center-of-mass contaminated states
by analyzing the dependence of coupled-cluster energies on
$\beta_{\rm CoM}$. As shown in Refs. \cite{epja2005,jpg2005},
the physical states obtained in coupled-cluster calculations are virtually
independent of $\beta_{\rm CoM}$, while the center-of-mass contaminated
states show a strong, nearly linear
dependence of excitation energies on $\beta_{\rm CoM}$.

We are currently working on alternative approaches to the 
effective interaction. In one of these
alternatives, instead of relying on the $G$-matrix method, we will
construct the renormalized Hamiltonian for coupled-cluster
calculations with the help of the
Lee-Suzuki procedure \cite{leesuzuki1,leesuzuki2,zuker1}, exploited in 
the no-core shell-model
approach \cite{navratil02}. This procedure will eliminate the
starting-energy dependence from our calculations. In particular,
we will investigate the differences between
our $G$-matrix approach and the no-core approach based on the
Lee-Suzuki transformation in the forthcoming work. 
Both methods have the appealing feature
that the effective interaction is properly renormalized as 
the size of the
harmonic oscillator basis is increased, approaching the bare 
Hamiltonian in the
limit of an infinite single-particle basis set. 
We will also study coupled-cluster applications using the 
the $V_{\rm lowk}$ effective interaction approach \cite{vlowk}. 
In this method, 
one uses the cutoff
in momentum space to soften the nucleon-nucleon interaction. 
The Lee-Suzuki-based and $G$-matrix-based
no-core procedures rely on the harmonic oscillator basis cutoff only,
while the $V_{\rm lowk}$ procedure uses the momentum-space cutoff
in addition to the basis set cutoff. The three methods
will produce identical results in the 
limit of an infinite basis set and infinite momenta.

Once the one- and two-body matrix elements of the
center-of-mass-corrected renormalized Hamiltonian
$H$, Eq. (\ref{hfinal}), are determined, we solve the nuclear many-body problem
using coupled-cluster theory. In order to
construct the coupled-cluster equations for the closed-shell $A$-body
system and the related PA-EOMCC and PR-EOMCC equations for the
$(A+1)$- and $(A-1)$-body nuclei in the computationally most
efficient way, as dictated by the factorized form of these equations
discussed in Secs.~\ref{subsec:singlerefccsd},
\ref{subsec:eomccsd}, and the Appendix,
we sort the one- and two-body matrix elements of
$H$ according to the particle-hole ($p\mbox{-}h$) character of
the single-particle indices that label them prior to
the coupled-cluster work. For example, the two-body matrix elements
$v_{\alpha\beta}^{\gamma\delta}$ defining $H$, Eq. (\ref{hfinal}), are sorted out
into six groups corresponding to the following $p\mbox{-}h$
types of single-particle indices $\alpha, \beta, \gamma, \delta$: $hhhh$,
$hhhp$, $hhpp$, $hphp$, $hppp$, $pppp$. The same philosophy of sorting
matrix elements according to the $p\mbox{-}h$ character of the relevant
single-particle indices is applied to one- and two-body matrix elements of the
similarity-transformed Hamiltonian of coupled-cluster theory discussed in the
next subsection.
This is a common practice in the most efficient
implementations of coupled-cluster methods by quantum chemists
and we follow the same recipe here.

\subsection{Brief Synopsis of the Single-Reference Coupled-Cluster Theory
and the Basic CCSD Approximation}
\label{subsec:singlerefccsd}

As mentioned earlier, the PA-EOMCC and PR-EOMCC calculations
for the $(A+1)$- and $(A-1)$-particle valence systems rely on the diagonalization of
the similarity-transformed Hamiltonian, obtained in the single-reference
coupled-cluster calculations for the $A$-particle closed-shell nucleus,
in appropriate subspaces of the Fock space.
Thus, before introducing the
PA- and PR-EOMCC methods for the $(A+1)$- and $(A-1)$-particle systems,
we provide the most essential information about the underlying closed-shell
coupled-cluster calculations that precede the PA- and PR-EOMCC steps.

The single-reference coupled-cluster theory
\cite{coester,coesterkummel,cizek,palduscizek}
is based on the exponential ansatz for the
ground-state wave function of the $A$-body system,
\begin{equation}
|\Psi_{0}^{(A)}\rangle = e^{T^{(A)}} |\Phi\rangle,
\label{eq:psi0}
\end{equation}
where $T^{(A)}$ is the cluster operator (a $p\mbox{-}h$ excitation operator)
and $|\Phi\rangle$ is the corresponding reference determinant
(defining the Fermi vacuum) obtained
by performing some mean-field calculation or by simply filling $A$
lowest-energy single-particle states (this is what we have done
in the calculations discussed in this paper). Here and elsewhere
in the present paper, we use superscripts, such as $(A)$, which indicate the
number of particles in a system under consideration, at the relevant
operators and energies. Normally, when $A$ remains fixed throughout the
entire calculation, this is
not essential, but in this paper we deal with systems with different
mass numbers (the
$A$-body as well as the $(A+1)$- and $(A-1)$-body systems), so that
it is useful to indicate the number of particles in a
many-body system of interest at the most essential mathematical quantities
to avoid confusion.

Formally,
Eq. (\ref{eq:psi0}) is a direct consequence of the
connected-cluster theorem, first clearly stated by Hubbard \cite{cct},
which is, in turn, related to the linked cluster theorem
of many-body perturbation theory \cite{cct,lct1,lct2,lct3}. According
to the connected-cluster theorem, the
cluster operator $T^{(A)}$ generates connected wave function diagrams
summed through infinite order.
Operationally, $T^{(A)}$ is a simple many-body excitation operator, which
in all standard coupled-cluster approximations is truncated at
a given (usually low)
$p\mbox{-}h$ excitation level $M < A$.
An example of the standard coupled-cluster
method is the CCSD (coupled-cluster singles and doubles)
approach \cite{purvis82,ccsdfritz,osaccsd},
which is used in this work to
obtain the ground-state information for the 
closed-shell ${^{16}}{\rm O}$ nucleus
We label coupled-cluster methods by
the standard acronyms adopted by chemists who
have led the development of coupled-cluster approaches for 
over 30 years now and
they are, by far, the most frequent users of
coupled-cluster approaches; 
see, for example, Refs. \cite{chem_rev2,kummel2002} for the relevant
historical remarks.
In this case, $M=2$ and the cluster operator $T^{(A)}$
is approximated by
\begin{equation}
T^{(A)}({\rm CCSD}) \equiv T^{(A)}(2) = T_{1} + T_{2} ,
\label{eq:t1t2}
\end{equation}
where
\begin{equation}
T_{1} = t_a^i a^{a} a_{i}
\label{eq:t1}
\end{equation}
and
\begin{equation}
T_2= \qua \, t_{ab}^{ij} a^{a} a^{b} a_{j} a_{i}
\label{eq:t2}
\end{equation}
are $1p\mbox{-}1h$ or singly excited and $2p\mbox{-}2h$ or
doubly excited cluster components,
$t_a^i$ and $t_{ab}^{ij}$ are the corresponding
singly and doubly excited cluster amplitudes, and
$i,j,\ldots$ ($a,b,\ldots$) are the single-particle
states occupied (unoccupied) in the reference determinant
$|\Phi\rangle$.
The general form of the truncated
cluster operator, defining a standard
single-reference coupled-cluster approximation characterized
by the excitation level $M$, is
\begin{equation}
T^{(A)}(M)=\sum\limits_{n=1}^{M} T_n ,
\label{a5}
\end{equation}
where
\begin{equation}
T_{n} = \left(\frac{1}{n!}\right)^{2}
t_{a_{1}\ldots a_{n}}^{i_{1} \ldots i_{n}} \,
a^{a_{1}} \cdots a^{a_{n}} a_{i_{n}} \cdots a_{i_{1}}
\label{eq:tn}
\end{equation}
($n = 1, \ldots, M$) are the many-body components
of $T^{(A)}(M)$, and $t_{a_{1}\ldots a_{n}}^{i_{1} \ldots i_{n}}$ are the
corresponding cluster amplitudes. 

The cluster amplitudes $t_{a_{1}\ldots a_{n}}^{i_{1} \ldots i_{n}}$
are determined by solving a coupled system of nonlinear and energy-independent
algebraic equations of the form:
\begin{equation}
\langle \Phi_{i_{1} \ldots i_{n}}^{a_{1} \ldots a_{n}} |
\bar{H}_{N}(M)|\Phi\rangle = 0 , \;\;\;\; i_{1}< \cdots < i_{n}, \;\;
a_{1} < \cdots < a_{n},
\label{ccaeq}
\end{equation}
where $n=1,\ldots,M$,
\begin{equation}
\bar{H}_{N}(M) = e^{-T^{(A)}(M)} H_{N} \, e^{T^{(A)}(M)}
= (H_{N} \, e^{T^{(A)}(M)})_{C}
\label{hbara}
\end{equation}
is the similarity-transformed
Hamiltonian of the coupled-cluster theory truncated at $Mp\mbox{-}Mh$
excitations, subscript $C$ designates
the connected part of the corresponding
operator expression, and $|\Phi_{i_{1} \ldots i_{n}}^{a_{1} \ldots a_{n}} \rangle
\equiv a^{a_{1}} \cdots a^{a_{n}} a_{i_{n}} \cdots a_{i_{1}}
|\Phi \rangle$ are the $np\mbox{-}nh$ or $n$-tuply
excited determinants relative to $|\Phi\rangle$.
The operator $H_{N}$ entering Eq. (\ref{hbara}) is the Hamiltonian
in the normal-ordered form relative to the $A$-particle Fermi vacuum
reference state $|\Phi\rangle$,
\beq
H_{N} =  H - \langle \Phi | H | \Phi \rangle = 
f_{\alpha}^{\beta} \, N[a^{\alpha} a_{\beta}]
+ \qua \, v_{\alpha\beta}^{\gamma\delta} \, N[a^{\alpha} a^{\beta} a_{\delta} a_{\gamma}] ,
\label{hnormal}
\eeq
where
$f_{\alpha}^{\beta} \equiv \langle \alpha | f | \beta \rangle
= z_{\alpha}^{\beta} + v_{\alpha i}^{\beta i}$
are matrix elements of the Fock matrix and $N[\cdots]$ designates the
normal product. 
In particular,
the standard CCSD equations for the singly and doubly
excited cluster amplitudes
$t_a^i$ and $t_{ab}^{ij}$, defining $T_1$ and $T_2$, respectively,
can be written as
\begin{equation}
\langle \Phi_{i}^{a} | \bar{H}_{N}{\rm (CCSD)}|\Phi\rangle = 0,
\label{ccsd1}
\end{equation}
\begin{equation}
\langle \Phi_{ij}^{ab} | \bar{H}_{N}{\rm (CCSD)}|\Phi\rangle = 0, \;\;\;
i < j, \; a < b ,
\label{ccsd2}
\end{equation}
where
\begin{eqnarray}
\bar{H}_{N}{\rm (CCSD)} & \equiv & \bar{H}_{N}(2) = e^{-T^{(A)}{\rm (CCSD)}}
H_{N} \, e^{T^{(A)}{\rm (CCSD)}}
\nonumber
\\
& = &
(H_{N} \, e^{T^{(A)}{\rm (CCSD)}})_{C}
\label{hccsd}
\end{eqnarray}
is the similarity-transformed Hamiltonian of the CCSD approach.
As mentioned in Sec. \ref{subsec:gmatrix}, we do not use the bare Hamiltonian in
our nuclear structure calculations.
Thus, the Hamiltonian $H$ used to construct the
similarity-transformed Hamiltonian $\bar{H}_{N}{\rm (CCSD)}$
for the nuclear structure calculations discussed in this paper
is replaced by the renormalized form of the Hamiltonian, Eq. (\ref{hfinal}),
resulting from the no-core $G$-matrix calculations.

The system of coupled-cluster equations, Eq.~(\ref{ccaeq}), is obtained
in the following way (suggested by {\v C}{\'\i}{\v z}ek
\cite{cizek}). We first insert
the coupled-cluster wave function $|\Psi_{0}^{(A)} \rangle$, Eq. (\ref{eq:psi0}),
into the $A$-body Schr{\" o}dinger equation,
\begin{equation}
H_{N} |\Psi_{0}^{(A)} \rangle = \Delta E_{0}^{(A)} |\Psi_{0}^{(A)} \rangle ,
\label{schreq}
\end{equation}
where
\beq
\Delta E_{0}^{(A)} = E_{0}^{(A)} - \langle \Phi | H | \Phi \rangle
\label{deltae}
\eeq
is the corresponding energy relative to the
reference energy $\langle \Phi | H | \Phi \rangle$,
and premultiply both sides of Eq. (\ref{schreq})
on the left by $e^{-T^{(A)}}$ to obtain
the connected-cluster form of the Schr{\" o}dinger equation
\cite{cizek,chem_rev1,chem_rev7,leszcz},
\begin{equation}
\bar{H}_{N} |\Phi \rangle = \Delta E_{0}^{(A)} |\Phi \rangle ,
\label{rightcc}
\end{equation}
where
\begin{equation}
\bar{H}_{N} = e^{-T^{(A)}} H_{N} \, e^{T^{(A)}}
= (H_{N} \, e^{T^{(A)}})_{C}
\label{similaritycc}
\end{equation}
is the similarity-transformed Hamiltonian.
Next, we project Eq. (\ref{rightcc}), in which
$T^{(A)}$ is replaced by its approximate form
$T^{(A)}(M)$, Eq. (\ref{a5}), onto the excited determinants
$|\Phi_{i_{1} \ldots i_{n}}^{a_{1} \ldots a_{n}} \rangle$, with
$n=1,\ldots,M$, corresponding
to the $p\mbox{-}h$ excitations included in $T^{(A)}(M)$.
The excited determinants
$|\Phi_{i_{1} \ldots i_{n}}^{a_{1} \ldots a_{n}} \rangle$ are orthogonal
to the reference determinant $|\Phi\rangle$, so that
we end up with the nonlinear and energy-independent
algebraic equations of the form of Eq. (\ref{ccaeq}).
Once the system of equations, Eq. (\ref{ccaeq}),
is solved for $T^{(A)}(M)$ or $t_{a_{1}\ldots a_{n}}^{i_{1} \ldots i_{n}}$
(or Eqs. (\ref{ccsd1}) and (\ref{ccsd2})
are solved for $T_{1}$ and $T_{2}$ or $t_a^i$ and $t_{ab}^{ij}$),
the ground-state coupled-cluster energy is calculated using the equation
\begin{eqnarray}
E_{0}^{(A)}(M) & = & \langle \Phi | H | \Phi \rangle + \Delta E_{0}^{(A)}(M)
\nonumber \\
& = &
\langle \Phi | H | \Phi \rangle
+ \langle\Phi|\bar{H}_{N}(M)|\Phi\rangle
\nonumber \\
& = &
\langle \Phi | H | \Phi \rangle +
\langle\Phi | \bar{H}_{N, {\rm close}}(M) |\Phi\rangle,
\label{egra}
\end{eqnarray}
where $\bar{H}_{N,{\rm close}}(M)$ is the closed part of
$\bar{H}_{N}{(M)}$ which is represented by those diagrams
contributing to $\bar{H}_{N}^{(M)}$ that have no external (uncontracted)
Fermion lines (as opposed to the open part of $\bar{H}_{N}(M)$
which is represented by the diagrams having
external or uncontracted Fermion lines; cf. Sec. \ref{subsec:eomccsd}).
It can easily be shown that if $H$
(in our case, the renormalized Hamiltonian defined by Eq. (\ref{hfinal})) does not contain
higher--than--two-body interactions and $2 \leq M \leq A$,
we can write
\beq
E_{0}^{(A)}(M) = \langle \Phi | H | \Phi \rangle +
\langle \Phi | [ H_{N} (T_{1} + T_{2} + \half T_{1}^{2}) ]_{C} | \Phi \rangle .
\label{egratwobody}
\eeq
In other words, we only need
$T_{1}$ and $T_{2}$ clusters to calculate the ground-state energy
$E_{0}^{(A)}(M)$ of the $A$-body ($A \geq 2$) system even if
we solve for other cluster components $T_{n}$ with $n > 2$.
Equation (\ref{egra}) can be
obtained by projecting the connected-cluster form of the Schr{\" o}dinger equation,
Eq. (\ref{rightcc}), on the reference configuration $|\Phi\rangle$
and replacing $T^{(A)}$ by $T^{(A)}(M)$.
In fact, the nonlinear character of the system of coupled-cluster
equations of the form of Eq. (\ref{ccaeq}) does not mean that
the resulting equations contain very high powers of $T^{(A)}(M)$.
For example, if the Hamiltonian $H$ (in our case, the
renormalized Hamiltonian obtained using the $G$-matrix technique) does not
contain higher--than--pairwise interactions,
the CCSD equations for the $T_{1}$ and $T_{2}$ clusters, or for the
amplitudes $t_a^i$ and $t_{ab}^{ij}$ that represent
these clusters, become
\begin{equation}
\langle \Phi_{i}^{a} |
[H_{N}(1 + T_{1} + T_{2} + \half T_{1}^{2} + T_{1} T_{2}
+ \six T_{1}^{3} )]_{C}
|\Phi \rangle = 0,
\label{momccsd1}
\end{equation}
\begin{eqnarray}
\langle \Phi_{ij}^{ab} | && \!\!\!\!\!\!\!
[H_{N}(1 + T_{1} + T_{2} + \half T_{1}^{2} + T_{1} T_{2}
+ \six T_{1}^{3}
\nonumber \\
&& + \half T_{2}^{2} + \half T_{1}^{2} T_{2}
+ \tfour T_{1}^{4} )]_{C}
|\Phi \rangle = 0.
\label{momccsd2}
\end{eqnarray}
In general, if the Hamiltonian $H$ contains two-body interactions only,
there are no terms in cluster components $T_{n}$ that are higher than
quartic terms in the coupled-cluster system, Eq. (\ref{ccaeq}),
independent of the truncation scheme $M$ used to define $T^{(A)}(M)$.
This is a purely mathematical statement, resulting from the fact that
we must connect external lines of $H_{N}$ with the vertices representing
the many-body components of the cluster operator $T^{(A)}$ to
determine the connected operator product represented
by $\bar{H}_{N}$, Eq. (\ref{similaritycc}),
and
not the result of some arbitrary truncation of the
exponential coupled-cluster wave function. 

In this work, we apply the CCSD approach to the closed-shell ${^{16}}$O nucleus
and, what is even more important in this particular study,
to obtain the corresponding similarity-transformed Hamiltonian
$\bar{H}_{N}{\rm (CCSD)}$, Eq. (\ref{hccsd}), which is used in the subsequent
PR-EOMCC and PA-EOMCC calculations for the ground and excited states
of the 15- and 17-particle nuclei.
As shown in our earlier papers \cite{dean04,kowalski04,wloch05,dean05,epja2005,jpg2005},
the CCSD method is perfectly adequate for a highly accurate description of the
ground state of ${^{16}}$O, providing the converged description of the ground state of
${^{16}}$O for the Hamiltonians containing two-body interactions
(the three-body clusters $T_{3}$ are negligible
\cite{kowalski04,wloch05,epja2005,jpg2005}). The highly accurate
description of the ground state of ${^{16}}$O is essential for obtaining a
well-balanced description of the 15- and 17-particle valence systems around
${^{16}}$O. The ability of the CCSD approach to
provide a highly accurate description of ${^{16}}$O is also useful from
the practical point of view. When properly implemented,
the CCSD approach is characterized
by the relatively inexpensive $n_{o}^{2}n_{u}^{4}$ steps, where
$n_{o}$ and $n_{u}$ are the numbers of
occupied and unoccupied orbitals, 
respectively, in the single-particle basis set,
making it applicable to systems with the larger values of $A$.
Our codes are already efficient enough to perform the CCSD
calculations with $A \sim 20-40$ and up to eight major oscillator shells,
which at least for $A=16$ is sufficient to obtain the converged
description \cite{wloch05}.

The explicit and computationally efficient
form of the CCSD and other coupled-cluster equations that can be used in routine
calculations for many-body systems with larger $A$ values and larger basis sets,
in terms of one- and two-body matrix elements of the
(renormalized form of the) Hamiltonian and cluster amplitudes
$t_{a_{1}\ldots a_{n}}^{i_{1} \ldots i_{n}}$ (in the CCSD case,
$t_a^i$ and $t_{ab}^{ij}$) can be most conveniently derived by applying
diagrammatic techniques of many-body theory combined with
diagram factorization methods which yield highly vectorized
computer codes \cite{piecuch2005,dean04,sak1,ccgamess,wloch2005}.
Once these equations are properly coded, we solve them using
iterative procedures, such as DIIS \cite{pulay1} 
(see Refs. \cite{ccgamess,ccdiis}).
The explicitly connected form of the coupled-cluster equations,
such as Eqs. (\ref{ccaeq}) or (\ref{momccsd1}) and (\ref{momccsd2}),
guarantees that the process of solving these equations
leads to connected terms in cluster components $T_{n}$ and
connected terms in the energy $E_{0}^{(A)}(M)$, independent of
the truncation scheme $M$ used to define $T^{(A)}(M)$.
The absence of disconnected terms in $T^{(A)}(M)$ and $E_{0}^{(A)}(M)$
is essential to obtain the rigorously size-extensive results.
The computationally efficient form
of the CCSD equations for the case of pairwise interactions,
which can be applied in large-scale nuclear structure
calculations such as those discussed in this work,
in terms of one- and two-body
matrix elements of the similarity-transformed Hamiltonian
$\bar{H}_{N}({\rm CCSD})$ that serve as the most natural intermediates
for setting up these equations and other recursively generated
intermediate quantities that are used to obtain a perfectly vectorized computer code,
is given in the Appendix.

\subsection{Equation-of-Motion Coupled-Cluster Methods for Valence Systems:
The PA-EOMCCSD and PR-EOMCCSD Approximations and their Implementation for
Nuclear Structure Calculations}
\label{subsec:eomccsd}

In addition to providing natural intermediates for setting up
coupled-cluster equations,
the use of the similarity-transformed Hamiltonians, 
$\bar{H}_{N}{(M)}$ or $\bar{H}_{N}{\rm (CCSD)}$, Eqs. (\ref{hbara})
or (\ref{hccsd}), respectively, in coupled-cluster calculations
provides a natural mechanism for extending
the ground-state coupled-cluster theory
to excited states of a given
$A$-body system or to
ground and excited states of the $(A+k)$- or $(A-k)$-particle
systems obtained by attaching $k$ particles to or removing $k$ particles
from the $A$-particle closed-shell core. This can be most
efficiently done by exploiting the EOMCC formalism
\cite{oureom,eomcc1,eomcc3} and its
PA-EOMCC (in chemistry, EA-EOMCC \cite{oureom,eaccsd1,eaccsd2,eaccsdt1,jeffeaip}) and
PR-EOMCC (in chemistry, IP-EOMCC \cite{chem_rev3,oureom,jeffeaip,%
ipccsd2,ipccsd3,ipccsd4,ipccsdt1,ipccsdt2,ipccsdt-3}) variants, and their
various multiply attached and multiply removed or ionized
(cf., e.g., Refs. \cite{jeffeaip,dipea1,dipea2,dipea3}) extensions
(see, also, Refs. \cite{lrcc0,lrcc1,lrcc2,lrcc3,lrcc4}
and \cite{sacci1,sacci2,sacci3,sacci4,sacci5,sacciopen}
for the related linear response and
symmetry-adapted cluster configuration interaction formalisms, respectively).
In all of these methods, we obtain excited states $|\Psi_{\mu}^{(A)}\rangle$
($\mu > 0$) of the $A$-particle system
or ground and excited states $|\Psi_{\mu}^{(A \pm k)}\rangle$
($\mu \geq 0$) of the
$A \pm k$-particle ($k > 0$) systems by applying the suitably defined
excitation ($R_{\mu}^{(A)}$) or particle-attaching/particle-removing
($R_{\mu}^{(A \pm k)}$) operator to the
ground state obtained in the single-reference
coupled-cluster calculations for the closed-shell $A$-body system.
Operators $R_{\mu}^{(A)}$ and $R_{\mu}^{(A \pm k)}$ are obtained
by diagonalizing the similarity-transformed Hamiltonians, such as
$\bar{H}_{N}{(M)}$, Eq. (\ref{hbara}), in the case coupled-cluster theory truncated
at $M$-tuple excitations, or $\bar{H}_{N}{\rm (CCSD)}$, Eq. (\ref{hccsd}),
in the CCSD case, in the relevant $A$-particle and $(A \pm k)$-particle
subspaces of the Fock space. For example, in the EOMCCSD approximation \cite{eomcc1,eomcc3},
which is a basic EOMCC approximation for the calculations
of excited states of the $A$-particle system,
we represent excited states $|\Psi_{\mu}^{(A)}\rangle$ as
\beq
|\Psi_{\mu}^{(A)}\rangle = R_{\mu}^{(A)} |\Psi_{0}^{(A)}\rangle =
R_{\mu}^{(A)} e^{T^{(A)}} |\Phi\rangle
\label{eomccwf}
\eeq
and replace $T^{(A)}$ by
the cluster operator $T^{(A)}({\rm CCSD})$, Eq. (\ref{eq:t1t2}),
obtained in the CCSD calculations,
and $R_{\mu}^{(A)}$ by
\beq
R_{\mu}^{(A)}{\rm (CCSD)} \equiv R_{\mu}^{(A)}(2) = R_{\mu,0} + R_{\mu,1} + R_{\mu,2},
\label{eq:r1r2}
\end{equation}
where
\begin{equation}
R_{\mu,0} = r_{0} \: {\mathbf 1},
\label{eq:r0}
\end{equation}
\begin{equation}
R_{\mu,1} = r_a^i \: a^{a} a_{i} ,
\label{eq:r1}
\end{equation}
and
\begin{equation}
R_{\mu,2} = \qua \, r_{ab}^{ij} \: a^{a} a^{b} a_{j} a_{i}
\label{eq:r2}
\end{equation}
are the reference, $1p\mbox{-}1h$, and $2p\mbox{-}2h$ components of
$R_{\mu}^{(A)}{\rm (CCSD)}$, and $r_{0}$, $r_a^i$, and $r_{ab}^{ij}$
are the corresponding excitation amplitudes
(${\bf 1}$ in Eq. (\ref{eq:r0}) is a unit operator).
In a more general case of coupled-cluster theory truncated at $M$-tuple excitations,
where $T^{(A)}$ is approximated by $T^{(A)}(M)$,
Eq. (\ref{a5}), the corresponding $p\mbox{-}h$ excitation operator $R_{\mu}^{(A)}$
in Eq. (\ref{eomccwf}) is approximated by \cite{oureom}
\begin{equation}
R_{\mu}^{(A)}(M)= \sum_{n=0}^{M} R_{\mu,n} ,
\label{ra}
\end{equation}
where, in analogy to $T_{n}$,
\begin{equation}
R_{\mu,n} =
\left(\frac{1}{n!}\right)^{2}
r_{a_{1}\ldots a_{n}}^{i_{1} \ldots i_{n}} \:
a^{a_{1}} \cdots a^{a_{n}} a_{i_{n}} \cdots a_{i_{1}} ,
\label{eq:rn}
\end{equation}
for $n \geq 1$, and $R_{\mu,0}$ is defined by Eq. (\ref{eq:r0}).
In the EOMCC methods for excited states, sometimes referred to as the
EE-EOMCC (excitation energy EOMCC) approaches \cite{oureom},
the $p\mbox{-}h$ excitation amplitudes $r_{a_{1}\ldots a_{n}}^{i_{1} \ldots i_{n}}$
defining $R_{\mu}^{(A)}(M)$ and the corresponding excitation energies
$\omega_{\mu}^{(A)}(M) = E_{\mu}^{(A)}(M) - E_{0}^{(A)}(M)$ of the $A$-body system of interest
are obtained by diagonalizing the similarity-transformed Hamiltonian
$\bar{H}_{N}{(M)}$ or, more precisely, its open part
\begin{eqnarray}
\bH{}_{N,{\rm open}}(M) & = & ( H_{N} e^{T^{(A)}(M)} )_{C,{\rm open}}
\nonumber
\\
& = & e^{-T^{(A)}(M)} H_{N} \, e^{T^{(A)}(M)} - \bH{}_{N,{\rm close}}(M)
\nonumber
\\
& = & e^{-T^{(A)}(M)} H_{N} \, e^{T^{(A)}(M)} - \Delta E_{0}^{(A)}(M),
\nonumber
\\
\label{hbaropen}
\end{eqnarray}
which has at least two external Fermion lines,
in the subspace of the $A$-particle Hilbert space spanned by the same excited determinants
$|\Phi_{i_{1} \ldots i_{n}}^{a_{1} \ldots a_{n}} \rangle$ with $n=1,\ldots,M$ that
are used to solve the underlying ground-state coupled-cluster calculations
($\Delta E_{0}^{(A)}(M)$ in Eq. (\ref{hbaropen}) is the ground-state coupled-cluster
energy relative to the reference energy $\langle \Phi | H | \Phi \rangle$;
cf. Eq. (\ref{deltae})).
In particular, the $r_a^i$ and $r_{ab}^{ij}$
amplitudes of the standard EOMCCSD (or EE-EOMCCSD) theory and the corresponding
excitation energies $\omega_{\mu}^{(A)}{\rm (CCSD)}$ of the $A$-body system are obtained
by diagonalizing the open part of the similarity-transformed Hamiltonian
of the CCSD approach,
\beq
\bH{}_{N,{\rm open}}({\rm CCSD}) = ( H_{N} e^{T^{(A)}({\rm CCSD})} )_{C,{\rm open}} ,
\label{hccsdopen}
\eeq
in the subspace spanned by the singly and doubly
excited determinants $|\Phi_{i}^{a}\rangle$ and $|\Phi_{ij}^{ab}\rangle$
used to set up and solve the ground-state CCSD equations.
We used the EOMCCSD approach to calculate the selected excited states
of the ${^{4}}$He and ${^{16}}$O nuclei in Refs. \cite{kowalski04,wloch05,dean05,epja2005,jpg2005}.

The idea of diagonalizing the similarity-transformed Hamiltonians
$\bH{}_{N,{\rm open}}(M)$, Eq. (\ref{hbaropen}),
or $\bH{}_{N,{\rm open}}({\rm CCSD})$, Eq. (\ref{hccsdopen}), can be extended
to ground and excited states of open-shell nuclei with
$(A \pm k)$ particles by replacing the particle-conserving $p\mbox{-}h$
excitation operator $R_{\mu}^{(A)}$ in Eq. (\ref{eomccwf})
by the suitably defined particle-attaching or particle-removing operator
$R_{\mu}^{(A \pm k)}$. For example, in the
basic PA-EOMCCSD (in chemistry, EA-EOMCCSD) \cite{eaccsd1,eaccsd2}
and PR-EOMCCSD (in chemistry, IP-EOMCCSD) \cite{chem_rev3,ipccsd2,ipccsd3,ipccsd4}
approaches (cf., also, Ref. \onlinecite{oureom,jeffeaip}) exploited in this
work, we define the
wave functions of the $(A+1)$- and $(A-1)$-particle systems, respectively, as
\beq
|\Psi_{\mu}^{(A \pm 1)}\rangle = R_{\mu}^{(A \pm 1)} e^{T^{(A)}} |\Phi\rangle ,
\label{eaip}
\eeq
where $T^{(A)}$ is approximated by
$T^{(A)}({\rm CCSD})$, Eq. (\ref{eq:t1t2}),
obtained in the CCSD calculations for the $A$-particle closed-shell system,
and $R_{\mu}^{(A + 1)}$ and $R_{\mu}^{(A - 1)}$ are replaced by the appropriately truncated
operators,
\beq
R_{\mu}^{(A + 1)}(2p\mbox{-}1h)
= R_{\mu,1p} + R_{\mu,2p\mbox{-}1h} = r_{a} a^{a} + \half r_{ab}^{\;\, j} a^{a} a^{b} a_{j}
\label{r2p1h}
\eeq
and
\beq
R_{\mu}^{(A - 1)}(2h\mbox{-}1p)
= R_{\mu,1h} + R_{\mu,2h\mbox{-}1p} = r^{i} a_{i} + \half r_{\;\, b}^{ij} a^{b} a_{j} a_{i} ,
\label{r2h1p}
\eeq
respectively, which generate the $(A+1)$- and $(A-1)$-particle states from the $A$-particle
CCSD wave function $e^{T_{1}+T_{2}} |\Phi\rangle$.
The $1p$ and
$2p\mbox{-}1h$ amplitudes $r_{a}$ and $r_{ab}^{\;\:j}$, respectively,
entering Eq. (\ref{r2p1h}) and defining the PA-EOMCCSD model, and the $1h$ and
$2h\mbox{-}1p$ amplitudes $r^{i}$ and $r_{\;\, b}^{ij}$, respectively,
entering Eq. (\ref{r2h1p}) and defining the PR-EOMCCSD model,
are determined by solving the eigenvalue problem
\beq
(\bH{}_{N, {\rm open}} \, R_{\mu}^{(A \pm 1)})_{C} |\Phi \rangle =
\omega_{\mu}^{(A \pm 1)} R_{\mu}^{(A \pm 1)}
|\Phi \rangle ,
\label{eomcc}
\eeq
where $\bH{}_{N, {\rm open}}$ is replaced by
the similarity-transformed Hamiltonian of the CCSD theory,
$\bH{}_{N,{\rm open}}({\rm CCSD})$, Eq. (\ref{hccsdopen}),
in the relevant subspaces of the
$(A+1)$- and $(A-1)$-particle subspaces, ${\mathscr H}^{(A+1)}$
and ${\mathscr H}^{(A-1)}$, respectively, of the Fock space.
The subspace of ${\mathscr H}^{(A+1)}$
used to solve the PA-EOMCCSD eigenvalue problem is spanned by the
$|\Phi^{a}\rangle = a^{a} |\Phi\rangle$ and $|\Phi_{\;\: j}^{ab}\rangle
= a^{a} a^{b} a_{j} |\Phi\rangle$ determinants.
The subspace of ${\mathscr H}^{(A-1)}$ used to solve the
PR-EOMCCSD problem is spanned by the
$|\Phi_{i}\rangle = a_{i} |\Phi\rangle$ and $|\Phi_{ij}^{\;\: b}\rangle
= a^{b} a_{j} a_{i} |\Phi\rangle$ determinants. By solving Eq. (\ref{eomcc}), we directly
obtain the energy differences,
$\omega_{\mu}^{(A + 1)} = E_{\mu}^{(A+1)} - E_{0}^{(A)}$ in the
PA-EOMCCSD case, and $\omega_{\mu}^{(A - 1)} = E_{\mu}^{(A-1)} - E_{0}^{(A)}$ in the
PR-EOMCCSD case, where $E_{\mu}^{(A+1)}$ and $E_{\mu}^{(A-1)}$ are the energies of
ground ($\mu=0$) and excited ($\mu > 0$) states of the $(A+1)$- and $(A-1)$-particle
systems, respectively, and $E_{0}^{(A)}$ is the ground-state
coupled-cluster (in this case, CCSD) energy of the $A$-particle reference system.

In analogy to the ground-state coupled-cluster theory,
one can extend the PA-EOMCCSD and PR-EOMCCSD schemes to higher excitations.
For example, in the nuclear analogs of the quantum-chemical
EA-EOMCCSDT \cite{eaccsdt1,jeffeaip}
and IP-EOMCCSDT \cite{jeffeaip,ipccsdt1,ipccsdt2} methods, which we would call here
the PA-EOMCCSDT (PA-EOMCC singles, doubles, and triples) and
PR-EOMCCSDT (PR-EOMCC singles, doubles, and triples) approaches,
one truncates the ground-state
cluster operator $T^{(A)}$ at the $3p\mbox{-}3h$ clusters $T_{3}$ (the $M=3$
or CCSDT case)
and augments the $R_{\mu}^{(A + 1)}(2p\mbox{-}1h)$ and $R_{\mu}^{(A - 1)}(2h\mbox{-}1p)$
operators, Eqs. (\ref{r2p1h}) and (\ref{r2h1p}), respectively,
by the $3p\mbox{-}2h$ component
\beq
R_{\mu,3p\mbox{-}2h} = \twelve r_{abc}^{\;\, jk} a^{a} a^{b} a^{c} a_{k} a_{j} ,
\label{3p2h}
\eeq
in the $R_{\mu}^{(A + 1)}$ case, and the $3h\mbox{-}2p$ component
\beq
R_{\mu,3h\mbox{-}2p} = \twelve r_{\;\, bc}^{ijk} a^{b} a^{c} a_{k} a_{j} a_{i}
\label{3h2p}
\eeq
in the $R_{\mu}^{(A - 1)}$ case. The resulting PA-EOMCCSDT and PR-EOMCCSDT operators
$R_{\mu}^{(A + 1)}$ and $R_{\mu}^{(A - 1)}$ are
\beq
R_{\mu}^{(A + 1)}(3p\mbox{-}2h) = R_{\mu,1p} + R_{\mu,2p\mbox{-}1h} + R_{\mu,3p\mbox{-}2h}
\label{r3p2h}
\eeq
and
\beq
R_{\mu}^{(A - 1)}(3h\mbox{-}2p) = R_{\mu,1h} + R_{\mu,2h\mbox{-}1p} + R_{\mu,3h\mbox{-}2p} ,
\label{r3h2p}
\eeq
respectively.
In general, if the cluster operator $T^{(A)}$ defining the ground-state of the
$A$-body system is truncated at the $Mp\mbox{-}Mh$ component, as shown in Eq.
(\ref{a5}), one typically truncates the corresponding PA-EOMCC operator $R_{\mu}^{(A + 1)}$ and
the corresponding PR-EOMCC operator $R_{\mu}^{(A - 1)}$ as follows:
\beq
R_{\mu}^{(A + 1)} =
\sum_{n=0}^{M-1} R_{\mu,(n+1)p\mbox{-}nh} ,
\label{eq:rea}
\eeq
where
\begin{eqnarray}
R_{\mu,(n+1)p\mbox{-}nh} & = & \frac{1}{n! (n+1)!} \,
r_{a a_{1} \ldots a_{n}}^{\;\; i_{1} \ldots i_{n}}
\nonumber
\\
& & \times \,
a^{a} a^{a_{1}} \cdots a^{a_{n}} a_{i_{n}} \cdots a_{i_{1}}
\label{eq:reamany}
\end{eqnarray}
and
\beq
R_{\mu}^{(A - 1)} =
\sum_{n=0}^{M-1} R_{\mu,(n+1)h\mbox{-}np} ,
\label{eq:rip}
\eeq
where
\begin{eqnarray}
R_{\mu,(n+1)h\mbox{-}np} & = & \frac{1}{n! (n+1)!} \,
r_{\: a_{1} \ldots a_{n}}^{i i_{1} \ldots i_{n}}
\nonumber
\\
& & \times \,
a^{a_{1}} \cdots a^{a_{n}} a_{i_{n}} \cdots a_{i_{1}} a_{i} ,
\label{eq:ripmany}
\end{eqnarray}
although, at least in principle, one could extend the summations over $n$
in Eqs. (\ref{eq:rea}) and (\ref{eq:rip}) to
$\sum_{n=0}^{M}$ without affecting the explicit
connectedness of the left-hand side of the corresponding
PA-EOMCC and PR-EOMCC eigenvalue problems, Eq. (\ref{eomcc}).
The detailed discussion of the relationships
between truncation schemes in the $R_{\mu}^{(A \pm 1)}$
and $T^{(A)}$ operators in the PA-EOMCC or EA-EOMCC
and PR-EOMCC or IP-EOMCC calculations can be found in Ref. \cite{oureom}
(cf., also, Refs.
\cite{jeffeaip,hirataeaip} for additional comments and numerical tests).
The $(n+1)p\mbox{-}nh$ and $(n+1)h\mbox{-}np$
amplitudes $r_{a a_{1} \ldots a_{n}}^{\;\; i_{1} \ldots i_{n}}$ and
$r_{\: a_{1} \ldots a_{n}}^{i i_{1} \ldots i_{n}}$, entering the
$R_{\mu}^{(A + 1)}$ and $R_{\mu}^{(A - 1)}$ operators of Eqs. (\ref{eq:rea}) and
(\ref{eq:rip}), respectively, are obtained by solving the
eigenvalue problem defined by Eq. (\ref{eomcc}), in which
$\bH{}_{N, {\rm open}}$ is replaced
by $\bar{H}_{N}{(M)}$, Eq. (\ref{hbara}).
The subspace of ${\mathscr H}^{(A+1)}$ relevant to the corresponding
truncated PA-EOMCC problem is spanned by the
$|\Phi^{a}\rangle$ and $|\Phi^{a a_{1} \ldots a_{n}}_{\;\; i_{1} \ldots i_{n}} \rangle
= a^{a} a^{a_{1}} \cdots a^{a_{n}} a_{i_{n}} \cdots a_{i_{1}} |\Phi\rangle$
($n = 1, \ldots , M-1$) determinants.
Similarly, the subspace of ${\mathscr H}^{(A-1)}$ relevant to the corresponding
truncated PR-EOMCC problem is spanned by the
$|\Phi_{i}\rangle$ and $|\Phi^{\:\: a_{1} \ldots a_{n}}_{i i_{1} \ldots i_{n}} \rangle
= a^{a_{1}} \cdots a^{a_{n}} a_{i_{n}} \cdots a_{i_{1}} a_{i} |\Phi\rangle$
($n = 1, \ldots , M-1$) determinants.

The PA-EOMCC and PR-EOMCC methods, as described above, and their extensions
to two or more valence particle or holes via multiply attached or
multiply ionized schemes \cite{jeffeaip,dipea1,dipea2,dipea3}
offer several advantages compared to the equally accurate, but
usually a lot more complicated, genuine
multi-reference coupled-cluster methods of either the valence-universal
\cite{vucc1,vucc2} or the Hilbert-space or state-universal \cite{succ}
type that are specifically designed to handle general classes of open-shell problems.
Although there has been significant progress in recent years in
the development of genuine multi-reference coupled-cluster theories
\cite{pal,succ1,succ2,succ3,succ4,succ5,succ6,xlin1,xlin2,xlin3,xlin3a,%
fsccsdt1,fsccsdt2},
multi-reference coupled-cluster calculations are often plagued by
intruder states, unphysical, singular, and multiple solutions, and mathematical difficulties
with the proper adaptation of the corresponding equations to symmetries of
the Hamiltonian if one aims at the general-purpose computer codes
(cf., e.g., Refs.
\cite{succ1,succA,succB,succC,succD,succE,succF,vuintruder1,vuintruder2} for further information).
Some of these issues are currently being addressed (cf., e.g.,
Refs. \cite{succ3,succ4,succ5,succ6,xlin1,xlin2,xlin3,xlin3a,fsccsdt2}), but
none of these problems are present in the
PA-EOMCC and PR-EOMCC calculations, which could be viewed as the physically motivated,
intruder-state free, state-selective
modifications of the powerful and elegant
valence-universal multi-reference coupled-cluster schemes
pioneered by Mukherjee and Lindgren \cite{vucc1,vucc2}.
In particular, the use of the
similarity-transformed Hamiltonian $\bH{}_{N,{\rm open}}$ obtained in the coupled-cluster
calculations for an $A$-particle closed-shell system, which
commutes with the symmetry operators of the original Hamiltonian $H$,
automatically guarantees that
the PA-EOMCC and PR-EOMCC wave functions defined by Eq. (\ref{eaip}),
obtained by diagonalizing $\bH{}_{N,{\rm open}}$ in the appropriate subspaces of
${\mathscr H}^{(A+1)}$ and ${\mathscr H}^{(A-1)}$, are
automatically adapted to the symmetries of the Hamiltonian $H$ and
one does not have to worry about the symmetry-contamination issues facing the
implementations of the open-shell coupled-cluster schemes employing the
unrestricted reference determinants.

Our calculations for the ground and low-lying excited states
of the 15- and 17-particle nuclei around ${^{16}}$O, reported
in this work, have been performed with the basic PA-EOMCCSD and PR-EOMCCSD methods,
in which the ground-state of ${^{16}}$O is represented by the
CCSD wave function $e^{T_{1}+T_{2}} |\Phi\rangle$ and
the nucleon-attaching and nucleon-removing operators 
$R_{\mu}^{(A + 1)}$ and $R_{\mu}^{(A - 1)}$ are defined by
Eqs. (\ref{r2p1h}) and (\ref{r2h1p}), respectively.
We can obtain an accurate description of the $(A+1)$-particle nuclei
$^{17}$O and $^{17}$F with the PA-EOMCCSD
method, in which we include the $1p$ and $2p\mbox{-}1h$ excitations from the ${^{16}}$O core
to form the 17-particle systems, since the ground-states and the low-lying excited states of the
$^{17}$O and $^{17}$F nuclei that we have singled
out in this work are essentially one-quasi-particle states, except, perhaps,
for the $(3/2)_{1}^+$ states of $^{17}$O and $^{17}$F, which are resonances. Similarly, we can study the
$(A-1)$-particle nuclei $^{15}$O and $^{15}$N with the basic
PR-EOMCCSD approach, in which we include the
$1h$ and $2h\mbox{-}1p$ excitations from the ${^{16}}$O closed-shell core, since the
low-lying states of these nuclei are expected to be dominated by one-quasi-hole states with respect
to the $A$-body reference $^{16}$O nucleus.
As discussed in Ref.~\cite{pp1993}, there is, for example, almost no
experimental evidence for the fragmentation
of the quasi-hole $p_{1/2}$ and $p_{3/2}$ states of $^{16}$O.
The fact that we use the $1p$ and $2p\mbox{-}1h$ excitations in the PA-EOMCCSD calculations
to form the $(A+1)$-body systems and the fact that we use the $1h$ and $2h\mbox{-}1p$
excitations in the PR-EOMCCSD calculations for the $(A-1)$-body systems
mean that we include many of the same correlations as Fujii {\em et al.} \cite{fujii2004,so84}.
Their approach is analogous to a hermitian coupled-cluster approach (see Ref.~\cite{so84}).
There are, however, differences between our PA-EOMCCSD and PR-EOMCCSD
calculations and the calculations reported by Fujii {\em et al.}
We use a biorthogonal EOMCC formalism, based on diagonalizing the non-hermitian similarity-transformed
Hamiltonian $\bar{H}_{N}{\rm (CCSD)}$, Eq. (\ref{hccsd}), obtained in highly
accurate CCSD calculations for the $A$-body closed-shell nucleus, which has been very successful
in quantum chemistry and molecular physics and which brings a lot of correlations
within basic truncation schemes, such as EOMCCSD, PA-EOMCCSD, and PR-EOMCCSD, through
the presence of high-order correlation terms in $\bar{H}_{N}({\rm CCSD})$. We also
differ in the definition of the model space, since Fujii {\em et al.} use a model space similar to
that used in the no-core shell-model calculations \cite{navratil02}, in which a ``triangular''
energy cutoff is applied to Slater determinants included in the diagonalization of the Hamiltonian,
in addition to the usual single-particle basis set cutoff.
Such a model space cannot be used in coupled-cluster calculations since it
violates the Pauli principle in the summations over the intermediate states that
emerge through products of many-body components of
the cluster operator $T^{(A)}$ in coupled-cluster equations. As mentioned earlier,
the use of a given truncation scheme for the cluster operator $T^{(A)}$
implies specific truncation schemes for the EOMCC operators, such
as $R_{\mu}^{(A + 1)}$ and $R_{\mu}^{(A - 1)}$. Thus,
we use all $1p$ and $2p\mbox{-}1h$ or $1h$ and $2h\mbox{-}1p$ excitations
in the PA-EOMCCSD and PR-EOMCCSD calculations and all
$1p\mbox{-}1h$ and $2p\mbox{-}2h$ cluster amplitudes $t_{a}^{i}$ and $t_{ab}^{ij}$
that are allowed by a given single-particle basis set, without imposing additional
energy cutoffs on the determinants that these excitations correspond to,
producing many additional and important correlations
that are outside model spaces used in the no-core shell-model calculations.

The use of the CCSD method for describing the correlated ground-state of the ${^{16}}$O reference
nucleus in the PR-EOMCCSD and PA-EOMCCSD calculations for
the 15- and 17-particle systems around ${^{16}}$O
is justified by the virtually perfect agreement of the
CCSD results and the results of the exact shell-model diagonalization of
the Hamiltonian in the same model space as used
in the CCSD calculations \cite{kowalski04,horoi:unpub}.
As shown in Refs. \cite{wloch05,epja2005,jpg2005} (cf., also, Ref.
\cite{horoi:unpub}), the three-body clusters
$T_{3}$ play a negligible role in the calculations of the binding energy
of ${^{16}}$O when the Hamiltonian includes pairwise interactions.
They also have almost no effect on
the lowest-energy $J^{\pi} = 3^{-}$ state of ${^{16}}$O, which is dominated
by the $1p\mbox{-}1h$ excitations, if the Hamiltonian contains the two-body interactions only
\cite{wloch05,epja2005,jpg2005}.
In this case, the basic EOMCCSD method, defined by Eq. (\ref{eomccwf}) in which
$T^{(A)}$ is replaced by the CCSD operator $T^{(A)}({\rm CCSD})$, Eq. (\ref{eq:t1t2}),
and $R_{\mu}^{(A)}$ is defined by Eq. (\ref{eq:r1r2}), provides the virtually
converged description. Interestingly enough, our PA-EOMCCSD and PR-EOMCCSD
results for the binding energies of $^{17}$O/$^{17}$F
and $^{15}$O/$^{15}$N, reported in this work, enable us
to explain most of the 6 MeV difference between the converged coupled-cluster
and experimental results for the lowest $J^{\pi} = 3^{-}$ state of ${^{16}}$O,
reported, for example, in Ref. \cite{wloch05},
which, according to our analysis, may largely be caused by the three-body interactions
neglected in our calculations that affect the relevant single-particle
energy spacings
(see Sec.~\ref{sec:resB} for a detailed discussion).

If we were to incorporate $T_{3}$ components in the ground-state coupled-cluster
calculations for ${^{16}}$O, we
would also have to include the $3p\mbox{-}2h$ excitations
$R_{\mu,3p\mbox{-}2h}$, Eq. (\ref{3p2h}), in the calculations for the 17-particle nuclei
and the $3h\mbox{-}2p$ excitations $R_{\mu,3h\mbox{-}2p}$, Eq. (\ref{3h2p}),
in the calculations for the 15-particle nuclei. Although we do not expect the
$R_{\mu,3p\mbox{-}2h}$ and $R_{\mu,3h\mbox{-}2p}$ terms to be significant for the
calculations discussed in this work, we will examine their role in the future
work, once the corresponding highly efficient computer codes for large-scale
nuclear applications are developed. As explained in Refs.
\cite{oureom,jeffeaip}, it is formally possible to include the
$R_{\mu,3p\mbox{-}2h}$ and $R_{\mu,3h\mbox{-}2p}$ components
in the PA-EOMCCSD and PR-EOMCCSD approaches,
which are then called the PA-EOMCCSD($3p\mbox{-}2h$) and PR-EOMCCSD($3h\mbox{-}2p$)
methods, respectively \cite{jeffeaip}, without losing the
explicitly connected form of the left-hand side of the corresponding
PA-EOMCC and PR-EOMCC eigenvalue problems given by Eq. (\ref{eomcc}).
A promising new development in the area of including
the relatively expensive $R_{\mu,3p\mbox{-}2h}$ and $R_{\mu,3h\mbox{-}2p}$ terms
in the PA-EOMCC and PR-EOMCC methods is that of the
so-called active-space PA-EOMCC (EA-EOMCC) and PR-EOMCC (IP-EOMCC) approaches, which
enable one to reduce the costs of the corresponding parent PA-EOMCCSDT (or EA-EOMCCSDT),
PR-EOMCCSDT (or IP-EOMCCSDT), and other more expensive higher-order
PA-EOMCC and PR-EOMCC calculations through a small subset of active
single-particle states, which in the nuclear physics context would correspond to
the highest occupied and lowest unoccupied shells of the
$A$-body reference nucleus (see Ref. \cite{jeffeaip}
for details). We will investigate the effects of the
$3p\mbox{-}2h$ and $3h\mbox{-}2p$ excitations on the PA-EOMCCSD and PR-EOMCCSD results
for $^{15}$O, $^{17}$O, $^{15}$N, and $^{17}$F reported in this study
using the PA-EOMCCSD($3p\mbox{-}2h$), PR-EOMCCSD($3h\mbox{-}2p$) approaches
and the active-space variants of the
PA-EOMCCSDT and PR-EOMCCSDT methods of Ref. \cite{jeffeaip} in
our future works.

We end this section by presenting the most essential algorithmic details
of the highly efficient PA-EOMCCSD and PR-EOMCCSD computer codes that can be applied to
valence systems with one valence particle
and one valence hole around the closed shell nucleus, such as $^{16}$O, and
other computational details pertinent to the specific calculations for the
$^{15}$O, $^{17}$O, $^{15}$N, and $^{17}$F nuclei discussed in this paper.
First, as discussed in Sec.~\ref{subsec:gmatrix},
we compute a $G$ matrix in a harmonic oscillator basis. In the case of the
calculations for the 15--17-particle systems described in this paper, we used model spaces
consisting of five to eight major oscillator shells; the $G$ matrix
and its pertinent two-body effective
interaction were computed for oscillator energies in the range $\hbar\Omega \in [10,20]$ MeV
and the optimum $\hbar\Omega$ value for each single-particle basis set was determined
by finding the minimum on the curve representing the dependence
of the CCSD ground-state energy of $^{16}$O on $\hbar\Omega$ (see Ref.~\cite{dean04}
for the details). The renormalized Hamiltonian resulting from the $G$ matrix calculations
was corrected for the spurious center-of-mass motion using Eq. (\ref{hfinal}),
in which $\beta_{\rm CoM}$ was chosen such that the expectation value of
the center-of-mass Hamiltonian $H_{\rm CoM}$ with the ground-state CCSD
wave function of $^{16}$O is 0.0 MeV. Since the PA-EOMCCSD and PR-EOMCCSD
methods are based on the linear-response-like
idea of directly calculating the energy differences
$\omega_{\mu}^{(A \pm 1)} = E_{\mu}^{(A \pm 1)} - E_{0}^{(A)}$
between the $(A \pm 1)$- and $A$-particle systems rather than the total
energies themselves, where
the ground-state of the closed-shell $A$-particle system serves as a reference
for the $(A \pm 1)$-body systems and
where we diagonalize the similarity-transformed Hamiltonian obtained in the
CCSD calculations for the $A$-particle reference system, we used the optimum values of $\hbar\Omega$
and $\beta_{\rm CoM}$ determined for $^{16}$O in the final
PA-EOMCCSD and PR-EOMCCSD calculations for $^{15}$O, $^{17}$O, $^{15}$N, and $^{17}$F
reported in this paper.
As already mentioned, for larger single-particle basis sets with seven or eight
major oscillator shells, the dependence
of the results on $\hbar\Omega$ is virtually none. The dependence of the
energies of physical states on $\beta_{\rm CoM}$ is virtually none as well
(cf. Refs. \cite{dean04,wloch05,dean05,epja2005,jpg2005} for the details).
Thus, the specific values of $\hbar\Omega$ and $\beta_{\rm CoM}$
become less and less important when the physical states are identified
and when larger basis sets are employed (in fact,
the optimum $\beta_{\rm CoM}$ values approach
zero as the basis set increases; cf. Table \ref{tab:be}).
Most of our calculations for $^{15}$O, $^{17}$O, $^{15}$N, and $^{17}$F were
based on the N$^3$LO nucleon-nucleon interaction model of Machleidt and
co-workers \cite{entem2003},
although we also performed the calculations for the CD-Bonn interaction model
\cite{cdbonn2000} and
the $V_{18}$ model of the Argonne group \cite{v18}. As pointed out earlier,
the Coulomb interaction was included in all of the calculations.

Once the one- and two-body matrix elements of the center-of-mass-corrected effective
Hamiltonian, Eq. (\ref{hfinal}), are determined and properly sorted out, as described
in Sec.~\ref{subsec:gmatrix}, we set up and solve the CCSD equations for the
closed-shell $A$-particle system (in our case, $^{16}$O), the PA-EOMCCSD equations
for the $(A+1)$-particle systems ($^{17}$O and $^{17}$F), and the PR-EOMCCSD equations
for the $(A-1)$-particle systems ($^{15}$O and $^{15}$N). Our ground-state CCSD computer codes
rely on the DIIS solver \cite{pulay1} (see, also, Refs. \cite{ccgamess,ccdiis}),
whereas the PA-EOMCCSD and PR-EOMCCSD equations for ground and excited states
of the $(A+1)$- and $(A-1)$-particle nuclei are solved with the
Hirao-Nakatsuji generalization \cite{hirao} of the Davidson
diagonalization algorithm \cite{dav} to nonhermitian eigenvalue
problems. The computationally efficient form of the CCSD
equations in terms of recursively generated intermediates
can be derived diagrammatically using Eqs. (\ref{momccsd1}) and (\ref{momccsd2}).
From the point of view of code efficiency, it is important to realize that some of the
intermediates entering the CCSD and other CCSD-based equations represent
matrix elements of the one- and two-body components of the
CCSD similarity-transformed Hamiltonian
$\bar{H}_{N,{\rm open}}({\rm CCSD})$, Eq. (\ref{hccsdopen}).
If $\bar{H}_{n}$ is the $n$-body
component of $\bar{H}_{N,\rm{open}}^{\rm{(CCSD)}}$,
for the one- and two-body components $\bar{H}_{1}$ and $\bar{H}_{2}$, respectively,
we can write
\beq
\bar{H}_{1} = \bar{h}_{\alpha}^{\beta} \, N[a^{\alpha} a_{\beta}]
\label{hbar1}
\eeq
and
\beq
\bar{H}_{2} = \qua \bar{h}_{\alpha\beta}^{\gamma\delta} \,
N[a^{\alpha} a^{\beta} a_{\delta} a_{\gamma}] ,
\label{hbar2}
\eeq
where $\bar{h}_{\alpha}^{\beta}$ and $\bar{h}_{\alpha\beta}^{\gamma\delta}$
are the one- and two-body matrix elements of $\bar{H}_{N,\rm{open}}^{\rm{(CCSD)}}$
that enter the CCSD, PA-EOMCCSD, and PR-EOMCCSD equations.
As shown in the Appendix, matrix elements
$\bar{h}_{\alpha}^{\beta}$ and $\bar{h}_{\alpha\beta}^{\gamma\delta}$
are calculated using the one- and two-body
matrix elements of the Hamiltonian in the normal-ordered form,
$f_{\alpha}^{\beta}$ and $v_{\alpha\beta}^{\gamma\delta}$, respectively
(cf. Eq. \ref{hnormal})) and the singly and doubly
excited cluster amplitudes
$t_a^i$ and $t_{ab}^{ij}$, defining $T_1$ and $T_2$, respectively.
The computationally efficient form
of the CCSD equations for the case of pairwise interactions in $H$,
in terms of selected types of $\bar{h}_{\alpha}^{\beta}$ and
$\bar{h}_{\alpha\beta}^{\gamma\delta}$ and other
recursively generated intermediates is given in the Appendix.

We also give in the Appendix the computationally
efficient form of the equations defining the PA-EOMCCSD
and PR-EOMCCSD eigenvalue problems. These are obtained
by applying diagrammatic methods to the PA-EOMCCSD and
PR-EOMCCSD equations, which can be given the following general form:
\begin{eqnarray}
\langle \Phi^{a} | \!\!\!\!\!\!\!\!\!\!\! && \lbrack (\bar{H}_{1} R_{\mu,1p})_{C}+
\sum_{n=1}^{2} (\bar{H}_{n} R_{\mu,2p \mbox{-}1h})_{C} \rbrack |\Phi\rangle
\nonumber \\
& = & \omega_{\mu}^{(A+1)} \, r_{a} ,
\label{eaeom1}
\\
\langle \Phi_{\hspace{4 pt} j}^{ab} | \!\!\!\!\!\!\!\!\!\!\! && \lbrack (\bar{H}_{2} R_{\mu,1p})_{C}+
\sum_{n=1}^{3} (\bar{H}_{n} R_{\mu,2p \mbox{-}1h})_{C} \rbrack |\Phi\rangle
\nonumber \\
& = & \omega_{\mu}^{(A+1)} \, r_{ab}^{\hspace{4 pt} j} ,
\label{eaeom2}
\end{eqnarray}
in the PA-EOMCCSD case, and
\begin{eqnarray}
\langle \Phi_{i} | \!\!\!\!\!\!\!\!\!\!\! && \lbrack (\bar{H}_{1} R_{\mu,1h})_{C}+
\sum_{n=1}^{2} (\bar{H}_{n} R_{\mu,2h \mbox{-}1p})_{C} \rbrack |\Phi\rangle
\nonumber \\
& = & \omega_{\mu}^{(A-1)} \, r^{i} ,
\label{ipeom1}
\\
\langle \Phi^{\hspace{3 pt} b}_{ij} | \!\!\!\!\!\!\!\!\!\!\! && \lbrack (\bar{H}_{2} R_{\mu,1h})_{C}+
\sum_{n=1}^{3} (\bar{H}_{n} R_{\mu,2h \mbox{-}1p})_{C} \rbrack |\Phi\rangle
\nonumber \\
& = & \omega_{\mu}^{(A-1)} \, r^{ij}_{\hspace{3 pt} b} ,
\label{ipeom2}
\end{eqnarray}
in the PR-EOMCCSD case. Although formally the PA-EOMCCSD and
PR-EOMCCSD equations require the consideration of the
three-body components of $\bar{H}_{N,\rm{open}}^{\rm{(CCSD)}}$ (cf.
the $n=3$ terms in Eqs. (\ref{eaeom2}) and (\ref{ipeom2})),
we do not have to calculate the corresponding six-index matrix elements
$\bar{h}_{\alpha\beta\gamma}^{\delta\epsilon\eta}$ explicitly. With the help of diagrammatic techniques,
the three-body components of $\bar{H}_{N,\rm{open}}^{\rm{(CCSD)}}$
that enter the PA-EOMCCSD and PR-EOMCCSD equations can be rigorously factorized
and rewritten in terms of the one- and two-body components of
$\bar{H}_{N,\rm{open}}^{\rm{(CCSD)}}$. In consequence, the final
working equations of the PA-EOMCCSD and PR-EOMCCSD methods
in terms of one- and two-body matrix elements of the Hamiltonian,
$f_{\alpha}^{\beta}$ and $v_{\alpha\beta}^{\gamma\delta}$, respectively,
$T_1$ and $T_2$ cluster amplitudes defining the underlying
$A$-particle ground-state CCSD problem, and the
$R_{\mu,1p}$, $R_{\mu,2p\mbox{-}1h}$,
$R_{\mu,1h}$, and $R_{\mu,2h\mbox{-}1p}$ excitation amplitudes defining the
particle-attaching and particle-removing operators,
$R_{\mu}^{(A+1)}(2p\mbox{-}1h)$ and $R_{\mu}^{(A-1)}(2h\mbox{-}1p)$, respectively,
can be re-expressed in terms of the one- and two-body
matrix elements of $\bar{H}_{N,\rm{open}}^{\rm{(CCSD)}}$,
$\bar{h}_{\alpha}^{\beta}$ and $\bar{h}_{\alpha\beta}^{\gamma\delta}$,
respectively, and a few additional recursively generated
intermediates, leading to a fully vectorizable algorithm.
This computationally efficient form of the
PA-EOMCCSD and PR-EOMCCSD equations is given in the Appendix too.

In addition to code vectorization,
the main advantage of deriving the CCSD, PA-EOMCCSD, and PR-EOMCCSD equations
in the form shown in the Appendix is the possibility of
obtaining the relatively low CPU operation count that characterizes these methods.
The CCSD equations and the determination of the full set of
one- and two-body matrix elements of $\bar{H}_{N,\rm{open}}^{\rm{(CCSD)}}$
are characterized by the $n_{o}^{2}n_{u}^{4}$ steps, where, as mentioned earlier,
$n_{o}$ and $n_{u}$ are the numbers of
occupied and unoccupied orbitals, respectively, in the single-particle basis set.
Once the CCSD equations are solved and
all one- and two-body matrix elements of $\bar{H}_{N,\rm{open}}^{\rm{(CCSD)}}$
are determined, the most expensive steps of the
PA-EOMCCSD and PR-EOMCCSD methods employing the factorized
equations shown in the Appendix are $n_{o}n_{u}^{4}$ and
$n_{o}^{2}n_{u}^{3}$, respectively. These relatively low, ${\mathscr N}^{5} - {\mathscr N}^{6}$
scalings of the costs of the CCSD, PA-EOMCCSD, and PR-EOMCCSD calculations with the
system size (${\mathscr N}$),
which are often orders of magnitude smaller than the costs of shell-model
calculations aimed at similar accuracies,
are among the most important advantages of the coupled-cluster
methodology pursued in this work.

\section{Results and Discussion}
\label{sec:results}

\subsection{Results of the PR-EOMCCSD and PA-EOMCCSD Calculations
with the N$^3$LO Interaction and their Convergence with the
Basis Set}
\label{sec:resA}

We focus first on the convergence of our PR-EOMCCSD and PA-EOMCCSD
results for the ground and excited states of $^{15}$O, $^{15}$N, $^{17}$O and
$^{17}$F with the size of the single-particle basis set used in
the coupled-cluster calculations
(see Tables \ref{tab:be} and \ref{tab:spectra}) and address the issue of
the dependence of these results on the choice of the oscillator parameter
$\hbar\Omega$. For comparison purposes,
we also list our previously published ground-state
CCSD results for $^{16}$O \cite{wloch05} (cf. Table \ref{tab:be}),
since $^{16}$O serves as a reference
nucleus for the PR-EOMCCSD and PA-EOMCCSD calculations.
We limit our discussion of the convergence properties of the
PR-EOMCCSD and PA-EOMCCSD results for $^{15}$O, $^{15}$N, $^{17}$O, and
$^{17}$F to the N$^3$LO interaction model \cite{entem2003}.
The PR-EOMCCSD and PA-EOMCCSD
results for the CD-Bonn \cite{cdbonn2000} and Argonne $V_{18}$ \cite{v18}
interactions exhibit almost identical qualitative features in terms of
their convergence with the number of major oscillator shells and the way they
depend on $\hbar\Omega$.
\begin{table}
\caption{Total binding energies and binding energies per particle
(in parentheses) for $^{15}$O and $^{15}$N (the PR-EOMCCSD
values), $^{16}$O (the CCSD values), and $^{17}$O and
$^{17}$F (the PA-EOMCCSD values),
computed with the N$^3$LO interaction model \cite{entem2003}, as functions
of the number of major oscillator shells $N$. All entries (except for
the unitless parameter $\beta_{\rm CoM}$)
are in MeV. The results for
$^{16}$O are taken from Ref.~\cite{wloch05}. The acronym Expt stands for the experimental
values, taken from Ref.~\cite{audi2003}. All energies were calculated
at the optimum values of $\hbar\Omega$ (the second last row;
determined by identifying the
$\hbar\Omega$ value at which the CCSD energy of $^{16}$O reaches the
minimum value) and $\beta_{\rm CoM}$ (the last row; determined by the condition
that the expectation value of $H_{\rm CoM}$ with the CCSD wave function
is 0.0 MeV). For eight major oscillator shells, the results are
virtually independent of $\hbar\Omega$ and $\beta_{\rm CoM} = 0.0$.
\label{tab:be}}
\begin{center}
\begin{tabular}{lrrrrr}
\hline
Nucleus & $N=5$ & $N=6$ & $N=7$ & $N=8$ & Expt \\
\hline
$^{15}$O &  91.047 &  93.772 &  93.219 &  92.376 & 111.955 \\
         &  (6.070)&  (6.251)&  (6.215)&  (6.158)&  (7.464)\\
$^{15}$N &  94.127 &  96.432 &  95.685 &  95.086 & 115.492 \\
         &  (6.275)&  (6.429)&  (6.379)&  (6.339)&  (7.699)\\
$^{16}$O & 108.943 & 113.341 & 112.446 & 111.221 & 127.619 \\
         &  (6.810)&  (7.084)&  (7.028)&  (6.951)&  (7.976)\\
$^{17}$O & 109.218 & 115.534 & 115.327 & 114.277 & 131.762 \\
         &  (6.425)&  (6.796)&  (6.784)&  (6.722)&  (7.751)\\
$^{17}$F & 106.097 & 112.869 & 112.782 & 111.510 & 128.220 \\
         &  (6.241)&  (6.640)&  (6.634)&  (6.559)&  (7.542)\\
$\hbar\Omega$ & 13 &      11 &      10 &      11 &         \\
$\beta_{\rm CoM}$ &1.50& 0.15 &   0.05 &     0.0 &         \\
\hline
\end{tabular}
\end{center}
\end{table}

As shown in Table \ref{tab:be}, the PR-EOMCCSD binding energies
of $^{15}$O and $^{15}$N, the results of the
CCSD calculations for the binding energy of
$^{16}$O, and the PA-EOMCCSD binding energies of $^{17}$O and
$^{17}$F are practically converged at the level of eight major oscillator shells.
As demonstrated earlier for $^{16}$O \cite{dean04},
the dependence of these results on the oscillator energy $\hbar\Omega \in [10,20]$ MeV
is almost negligible, particularly for seven or eight major oscillator shells.
In the latter case, the dependence of the results on $\hbar\Omega$ is practically none.
This tells us that the renormalization of the short-range part of the nucleon-nucleon
interaction with the no-core $G$-matrix approach combined with the inclusion of singly and doubly
excited clusters and the corresponding $1p$, $2p\mbox{-}1h$, $1h$, and
$2h\mbox{-}1p$ excitations in the coupled-cluster and PR-EOMCC/PA-EOMCC calculations for
the valence systems around $^{16}$O leads to reasonably well converged
ground-state energies of these systems. It is true that the N$^3$LO interaction model
has a rather soft core, since it carries a cutoff in relative momentum of $\Lambda = 500$ MeV.
Thus, in developing the effective two-body interaction based on N$^3$LO by diagonalizing the
deuteron in an oscillator basis, one obtains a converged result to six leading digits
with 50 to 60 oscillator shells for $\hbar\Omega \in [5,50]$ MeV.
For the CD-Bonn and $V_{18}$ interactions, one needs more than 
100 major shells in order to obtain a converged result for the deuteron. 
However, the advantage of the $G$-matrix approach used in this work is that we can renormalize
the short-range part of the interaction
exactly, since the free part of the $G$ matrix is computed
in a momentum basis first, with the relative momenta $|{\bf p}| \in [0,\infty)$. Thus, the renormalization
problems of the short-range part of the two-body interaction,
seen, for example, in the no-core approach \cite{navratil02}, with a relatively slow convergence
as a function of the harmonic oscillator excitations, are not present here. 
This means, in turn, that when we use this $G$ matrix in coupled-cluster calculations,
the results for all modern nucleon-nucleon potentials, such as 
N$^3$LO, CD-Bonn, and $V_{18}$ used here, are basically converged
within eight major shells. It should also be pointed out that
our results for N$^3$LO agree very well with those of Fujii {\em et al.} \cite{fujii2004}.

The fact that our $G$-matrix-based coupled-cluster results are essentially
converged with the basis set has several important consequences for nuclear
many-body theory. We can, for example, claim that any disagreements with
experimental data are, most likely, due to the missing degrees of freedom
in our Hamiltonians, such as three-nucleon interactions. Indeed, as shown in Table \ref{tab:be},
our coupled-cluster calculations miss the experimental binding energies
(taken from Ref.~\cite{audi2003}) by approximately $1.3-1.4$ MeV per nucleon for the $A=15$ nuclei
and by approximately $1$ MeV per nucleon for the $A=16$ and $A=17$ systems. At least
in principle, several factors can contribute to these differences, but
we believe that the three-body interactions are the primary source.
It is true, for example, that we are using the solution to a two-body problem
(our $G$ matrix) as the starting point for defining a many-body Hamiltonian
with pairwise interactions
for the $A=15-17$ nuclei,
and it is known that a two-body interaction derived from the diagonalization
of a three-body problem is different from the corresponding
two-body interaction derived by diagonalizing
the two-body problem (e.g., deuteron) \cite{navratil02,navratil1999,ellis2005}. However,
as the size of the model space is increased, both two-body interactions yield very similar
results (see the discussion in Ref.~\cite{navratil1999}).
Thus, since we use large model spaces with seven or even eight major oscillator shells,
the differences between these two types of effective two-body interactions are minimal
and cannot, as such, contribute to the differences between the coupled-cluster and
experimental data observed in our calculations. Clearly, we are
missing some correlations in our coupled-cluster calculations, which
ignore, for example, $T_{3}$ clusters in the ground-state
calculations for ${^{16}}$O and $3p\mbox{-}2h$
and $3h\mbox{-}2p$ components of $R_{\mu}^{(A + 1)}$ and $R_{\mu}^{(A - 1)}$
in the PR-EOMCC and PA-EOMCC calculations for the 15- and 17-particle nuclei,
but, as mentioned earlier, the ground and excited states of
$^{15}$O/$^{15}$N and $^{17}$O/$^{17}$F are essentially one-quasi-particle
and one-quasi-hole states, respectively, in which $3p\mbox{-}2h$
and $3h\mbox{-}2p$ excitations play a negligible role.
As shown in Refs. \cite{wloch05,epja2005,jpg2005},
$T_{3}$ clusters bring in at most a total of $1$ MeV in the ground-state calculations
for $^{16}$O (less than 0.1 MeV per nucleon) and cannot,
therefore, account for the observed differences
between the binding energies per nucleon. We can thus summarize this part of our discussion
by stating that the discrepancy between experiment and theory observed in
Table \ref{tab:be} can be ascribed to the
missing three-body interactions, which are not included in our effective Hamiltonians.
The advantage of the N$^3$LO model and similar models based on effective field theory, is that they allow
for a consistent derivation of three-body terms (see, for example, Refs.~\cite{entem2003,epelbaum2003}).
However, we have not developed the coupled-cluster codes for dealing with such interactions
yet. We plan to do it in the future work.

It is interesting to observe that although the PR-EOMCCSD/CCSD/PA-EOMCCSD approaches
underbind the five nuclei by about 1--1.4 MeV per particle, pointing to the need
for the incorporation of three-body forces, the relative binding energies
of $^{15}$O, $^{15}$N, $^{16}$O, $^{17}$O, and $^{17}$F
obtained in coupled-cluster calculations are in good agreement with experiment.
For example, the difference between experimental binding energies of
$^{16}$O and $^{17}$O is $0.225$ MeV per particle.
The CCSD and PA-EOMCCSD ground-state energies of $^{16}$O and $^{17}$O
resulting from the calculations with eight major oscillator shells
differ by 0.229 MeV per particle, in excellent agreement with experiment.
The differences between the binding energies for the $A=15$ nuclei and for the $A=17$ nuclei
are close to the experimental values too. From Table \ref{tab:be}, we extract
${\rm BE(^{15}N) - BE(^{15}O)} = 0.181$ MeV per particle,
when the PR-EOMCCSD/N$^3$LO approach with
eight major shells is employed, and 0.235 MeV per particle, when the experimental data are used
(${\rm BE = binding \; energy}$).
Similarly, the PA-EOMCCSD/N$^3$LO calculations with eight major shells give
${\rm BE(^{17}O) - BE(^{17}F)} = 0.163$ MeV per particle, which is in reasonable agreement
with the experimental result of 0.209 MeV per particle.
The binding energies per nucleon resulting from the
PR-EOMCCSD/CCSD/PA-EOMCCSD calculations
satisfy the ordering
$^{15}{\rm O} < \: ^{15}{\rm N} < \: ^{17}{\rm F} < \: ^{17}{\rm O} < \: ^{16}{\rm O}$.
With an exception of the $^{15}$N and $^{17}$F nuclei, whose binding
energies per particle are close to each other and ordered in experiment
as $^{17}{\rm F} < \: ^{15}{\rm N}$, the ordering resulting from
the PR-EOMCCSD/CCSD/PA-EOMCCSD calculations is correct.
This clearly illustrates that a great deal of useful information can be
extracted from the relatively inexpensive coupled-cluster
calculations of the CCSD type employing two-body interactions.
The total binding energies are affected by three-body forces. However, the relative binding
energies for the nuclei around $^{16}{\rm O}$ seem to be reasonably well described
at the two-body interaction level when the coupled-cluster methods are used
to describe particle correlations.

We end this subsection by tabulating the results of the PR-EOMCCSD and PA-EOMCCSD
calculations for the low-lying excited states of
$^{15}$O, $^{15}$N, $^{17}$O, and $^{17}$F obtained with the N$^3$LO potential
(see Table \ref{tab:spectra}; the experimental data are taken from Ref. \cite{fire}).
Except for the $(3/2)^+_1$ resonance states in 
$^{17}$O and $^{17}$F, all the other excited states listed
in Table \ref{tab:spectra} are expected to be strongly dominated by
one quasi-particle or quasi-hole states, meaning that the inclusion
of the $1p$ and $2p\mbox{-}1h$ excitations
in the PA-EOMCCSD calculations and the $1h$ and $2h\mbox{-}1p$ correlations in the PR-EOMCCSD calculations
should provide a reasonable description of these states. This is confirmed in Table \ref{tab:spectra}.
The PR-EOMCCSD/N$^3$LO results for the $(3/2)^-_1$ states of $^{15}$O and $^{15}$N,
employing seven or eight major oscillator shells, are
particularly impressive, producing errors relative to experiment that do not exceed 0.1 MeV.
For the particle case, the $^{17}$O $(1/2)^+_1$ excited state resulting from the
PA-EOMCCSD calculations is slightly below the $(5/2)^+_1$ ground state of
$^{17}$O, when the N$^3$LO interaction and eight oscillator shells are employed.
Else, the agreement with the experimental data is quite satisfactory.
Again, as in Table \ref{tab:be},
we note a reasonably good convergence in the PR-EOMCCSD and PA-EOMCCSD
results for the low-lying excited states
of $^{15}$O, $^{15}$N, $^{17}$O, and $^{17}$F
in terms of the number of harmonic oscillator shells in a basis, which is yet
another confirmation of the
effectiveness of our computational procedure, in which we perform coupled-cluster
calculations with the renormalized form of the Hamiltonian rather than with the
underlying bare interactions that would lead to a very slow convergence rate with the
number of single-particle states, making the calculations unmanageable.
\begin{table}
\caption{Energies of the
low-lying excited states of $^{15}$O, $^{15}$N, $^{17}$O and
$^{17}$F, relative to the corresponding ground-state energies
(the $(1/2)_1^-$ states of $^{15}$O and $^{15}$N and
the $(5/2)^+_1$ states of $^{17}$O and $^{17}$F),
computed with the N$^3$LO interaction model \cite{entem2003} and the
PR-EOMCCSD ($^{15}$O, $^{15}$N) and PA-EOMCCSD ($^{17}$O and $^{17}$F)
methods, as functions
of the number of major oscillator shells $N$. All entries are in MeV. Note that the
experimentally observed $(3/2)^+_1$ states in 
$^{17}$O and $^{17}$F are resonances. The experimental data (Expt) are from Ref.~\cite{fire}.
For the optimum values of $\hbar\Omega$ and $\beta_{\rm CoM}$, see
Table \ref{tab:be}.
\label{tab:spectra}}
\begin{center}
\begin{tabular}{lrrrrr}
\hline
Excited state & $N=5$ & $N=6$ & $N=7$ & $N=8$ & Expt \\
\hline
$^{15}$O $(3/2)_1^-$& 6.515 & 6.602 &  6.166 &  6.264 & 6.176 \\
$^{15}$N $(3/2)_1^-$& 6.354 & 6.680 &  6.256 &  6.318 & 6.323 \\
$^{17}$O $(3/2)^+_1$& 6.298 & 6.031 &  5.489 &  5.675 & 5.084 \\
$^{17}$O $(1/2)^+_1$& 0.328 & 0.130 & -0.349 & -0.025 & 0.870 \\
$^{17}$F $(3/2)^+_1$& 6.460 & 6.207 &  5.686 &  5.891 & 5.000 \\
$^{17}$F $(1/2)^+_1$& 0.748 & 0.544 &  0.088 &  0.428 & 0.495 \\
\hline
\end{tabular}
\end{center}
\end{table}

Within a single-particle picture, the splitting between the $(3/2)_1^-$ excited
and $(1/2)_1^-$ ground states in $^{15}$O and $^{15}$N and the splitting
between the $(3/2)^+_1$ excited and $(5/2)^+_1$ ground states in $^{17}$O and $^{17}$F
should arise from the nuclear spin-orbit force. It is interesting to analyze
to what extent the three-nucleon interactions may affect these splittings.
The nucleon-nucleon interaction contains a short-range spin-orbit force,
which in a meson-exchange model picture originates from heavier
vector mesons. Several partial waves receive significant contributions from the two-body spin-orbit force.
For example, the $^3P_2$ partial wave, crucial for the pairing properties in nuclei and neutron star matter,
yields an attractive interaction up to almost
1 GeV in laboratory energy for the two-nucleon scattering.
This attraction arises from the two-body spin-orbit force, since both the 
central and tensor force contributions are repulsive.
Within the framework of many-body perturbation theory,
the largest contribution to the spin-orbit force arises from the first-order Hartree-Fock diagram.
Indeed, for the N$3$LO model used here and for an oscillator energy $\hbar\Omega = 14$ MeV, we obtain an
excitation energy of $5.412$ MeV for the $0p_{3/2}$ state
of $^{15}$O, in reasonable agreement with the experimental and coupled-cluster data
in Table \ref{tab:spectra} 
\cite{comment}.
At the Hartree-Fock diagram level, the origin of the spin-orbit splitting comes then from the 
renormalization of the short range two-body spin-orbit force.  
The nuclear tensor force gives also, as a second- and higher-order 
process, a contribution to the single-particle spin-orbit splitting
(see the detailed discussion in Ref.~\cite{ab1981} for further information). The authors
of Ref.~\cite{ab1981} show how the 
second-order diagrams in many-body perturbation theory
with the $2h\mbox{-}1p$ and $2p\mbox{-}1h$ intermediate states
yield repulsive and attractive contributions to the single-particle energies, respectively.
Depending on the strength of the nuclear tensor force, the spin-orbit splittings can then be enhanced
or reduced. If the tensor force is weak, as is the case for the N$^3$LO model, the reduced higher-order
quenching of the tensor force terms enhances the spin-orbit splitting with respect to the Hartree-Fock 
diagram. Anticipating the discussion in Sec.~\ref{subsec:othermodel},
potentials with a stronger tensor force, such as the $V_{18}$ model of the Argonne group
\cite{v18}, result in a smaller spin-orbit splitting than the N$^3$LO model
(and a reduction in the $(3/2)_1^{-} -  (1/2)_1^-$ and
$(3/2)_1^{+} -  (5/2)_1^+$ spacings in the $^{15}$O/$^{15}$N and
$^{17}$O/$^{17}$F nuclei, respectively).
The authors of
Refs.~\cite{pp1993,ab1981} demonstrated then that a two-pion
three-nucleon interaction also contributes to the 
spin-orbit splitting. With the inclusion of such a term,
Pieper and Pandharipande \cite{pp1993}, reproduced very well  
the $(3/2)_1^{-} -  (1/2)_1^-$ splitting
in $^{15}$N. These findings were later corroborated by Heisenberg and Mihaila in their coupled-cluster
calculations with three-body interactions for $^{16}$O (see
Refs.~\cite{bogdan1,bogdan2,bogdan3,bogdan4} and the 
discussion in the next subsection as well). The fact that we reproduce very well
the experimental $(3/2)_1^{-} -  (1/2)_1^-$
spin-orbit splittings in $^{15}$N and $^{15}$O
with the pairwise N$^3$LO model indicates that the spin-orbit force associated with an
eventual three-body force for N$^3$LO should be small. This may be an important
finding for our understanding of the role of three-body forces in
nuclear structure calculations.

In Sec.~\ref{subsec:othermodel},
we present results for the binding energies and spectra of the
$^{15}$N and $^{15}$O nuclei and their $^{17}$F and $^{17}$O
counterparts using the
CD-Bonn \cite{cdbonn2000} and the Argonne $V_{18}$ \cite{v18} interaction models as well,
so that we can see how much the effects due to three-body interactions may depend
on the underlying two-body forces.
However, before we proceed, let us discuss interesting consequences of our
PR-EOMCCSD and PA-EOMCCSD calculations for $^{15}$N, $^{15}$O, $^{17}$F, and $^{17}$O
for the nuclear structure studies of the excited states of $^{16}$O.

\subsection{Consequences of the PR-EOMCCSD and PA-EOMCCSD
Calculations for the Valence Systems around $^{16}$O
for the Studies of Excitations in $^{16}$O}
\label{sec:resB}

Based on the N$^3$LO results discussed in the previous subsection,
we attempt to link our findings
to nuclear structure studies of the excitations in $^{16}$O.
The fact that we obtain practically converged results for a given 
two-body Hamiltonian allows us to infer that eventual disagreements
with experiment in the results of {\em ab initio} calculations
for excited states of $^{16}$O can very likely be retraced
to the degrees of freedom that are not included in the existing two-body Hamiltonians.

Here, we discuss the excited states of $^{16}$O with an expected $1p\mbox{-}1h$ structure.
In Ref.~\cite{wloch05}, we obtained converged results for the lowest-lying
$3^-_1$ state of $^{16}$O. For the N$^3$LO interaction,
we obtained an excitation energy of about 12 MeV, almost 6 MeV
above the experimental value of $6.13$ MeV. 
The same Hamiltonian as that used here was employed. The low-lying excited states of 
$^{16}$O and, in general, states which involve cross-shell excitations, have always eluded a proper
microscopic description (see, for example,
Refs.~\cite{andres1,pe70,pe71,hj90,wb92,wbm92} and references therein).

Let us concentrate on the lowest-energy
$3^-_1$ state of $^{16}$O. In a zero-order approximation, this state may be regarded as a state that
arises from the single $i \rightarrow a$ excitation
from the $i = 0p_{1/2}$ hole state to the $a = 0d_{5/2}$ particle state. Relative to the
$^{16}$O ground state, the energy required to produce such an excitation equals
\begin{eqnarray}
\Delta \epsilon_{\pi} & = & \epsilon_{\pi}(0d_{5/2}) - \epsilon_{\pi}(0p_{1/2})
\nonumber \\
& = &
[\mathrm{BE}(^{16}\mathrm{O})-\mathrm{BE}(^{17}\mathrm{F})]
+ [\mathrm{BE}(^{16}\mathrm{O})-\mathrm{BE}(^{15}\mathrm{N})]
\nonumber \\
& = & 11.526 \hspace{0.1cm}\mathrm{MeV},
\label{epi}
\end{eqnarray}
for the proton case, and
\begin{eqnarray}
\Delta \epsilon_{\nu} & = & \epsilon_{\nu}(0d_{5/2})-\epsilon_{\nu}(0p_{1/2})
\nonumber \\
& = &
[\mathrm{BE}(^{16}\mathrm{O})-\mathrm{BE}(^{17}\mathrm{O})]
+ [\mathrm{BE}(^{16}\mathrm{O})-\mathrm{BE}(^{15}\mathrm{O})]
\nonumber \\
& = & 11.521 \hspace{0.1cm}\mathrm{MeV},
\label{enu}
\end{eqnarray}
for the neutron case, where BE's in the above equations
represent the relevant total binding energies.
In calculating the above values of the $1p\mbox{-}1h$ excitation energies
$\Delta\epsilon_{\pi}$ and $\Delta\epsilon_{\nu}$ that provide us with the
zero-order estimates of the excitation energy of the lowest $3^-_1$ state of $^{16}$O,
we used the experimental binding energies listed in Table \ref{tab:be}.
As we can see from Eqs. (\ref{epi}) and (\ref{enu}),
the proton and neutron excitation energies are practically identical. This reflects
a well-known feature of the spin-isospin
saturated systems. Without interactions among nucleons
and with the above single-particle orbits used as the only active degrees
of freedom, all negative parity states with quantum numbers $J^{\pi}=2^-, \, 3^-$
would be at the above energies
of approximately 11.5 MeV. The interactions among nucleons lower the energy of the
first-excited $3^-$ state by $11.5-6.1=5.4$ MeV.

Let us now compare the
approximate energy spacing defining the lowest $3^-$ state of $^{16}\mathrm{O}$, resulting from the
use of experimental binding energies, as shown above ($11.5$ MeV), with the
values of $\Delta\epsilon_{\pi}$ and $\Delta\epsilon_{\nu}$ based on the
results of coupled-cluster calculations for the binding energies of
$^{16}\mathrm{O}$ and valence systems around $^{16}\mathrm{O}$ obtained with the N$3$LO interaction and
eight major oscillator shells. These results are $\Delta\epsilon_{\pi} = 15.846$ MeV
and $\Delta \epsilon_{\nu} = 15.789$ MeV, for proton and neutron excitations, respectively,
with almost the same difference between the proton and neutron cases as observed in experiment.
The authors of Ref.~\cite{fujii2004} obtained 14.72 MeV and 14.64 MeV for protons and neutrons, 
respectively, using the same N$^3$LO interaction as used here. We should note, however,
that their results for $^{16}$O are not fully converged as a function
of the model space size. Using the above elementary picture of the
$1p\mbox{-}1h$ excitation defining the lowest $3^-$ state of $^{16}\mathrm{O}$,
which involves only two orbits in the definition of the relevant model space, we can see
that we are off by approximately $15.8 - 11.5 = 4.3$ MeV, when we compare
the $\Delta\epsilon_{\pi}$ and $\Delta\epsilon_{\nu}$ energy spacings resulting from
coupled-cluster calculations with the experimental estimates of these spacings.
This difference is obviously an interaction and method dependent result. It is, however,
converged as a function of the number of oscillator shells in a basis set, showing that
the discrepancy of 4.3 MeV between theory an experiment for the energy gap 
between the $0p$ and $1s0d$ shells accounts for a large fraction
of the missing 6 MeV needed to reproduce the first $3^-$ state of $^{16}$O.
This is, perhaps, the most likely candidate for a consistent explanation of the
large difference between converged coupled-cluster result for the
lowest $3^-$ state of $^{16}$O and experiment reported in Ref. \cite{wloch05}.
The above analysis indicates that
a large fraction of the difference between theory and experiment can be traced in this case to
errors in reproducing the experimental binding energies of $^{16}$O and valence
systems around $^{16}$O by coupled-cluster methods employing pairwise interactions only.
This allows us to conclude that a 6 MeV difference between coupled-cluster result and experiment
for the lowest $3^-$ state of $^{16}$O is primarily caused by
the lack of three-body interactions in our calculations, and much less by the approximate
treatment of particle correlations by the coupled-cluster methods used in our studies.
The above analysis also implies
that with an adjusted gap between the $0p$ and $1s0d$ shells, one should be able to get
a better reproduction of the excited states of
$^{16}$O which have a well-defined $1p\mbox{-}1h$ structure, such as the
lowest $3^-$ state discussed here.
One possible strategy for describing excited states of closed-shell nuclei
dominated by $1p\mbox{-}1h$ excitations might be to
keep the original two-body Hamiltonian and add
additional three-body terms via corrections to the single-particle energies, 
as advocated recently by Zuker \cite{andres2,andres3}.

\subsection{Results for Other Interaction Models}
\label{subsec:othermodel}

The binding energies per particle for the three interaction models examined in
this work, namely N$^3$LO, CD-Bonn, and $V_{18}$, are listed in Table \ref{tab:becomp}.
We only show the essentially converged results obtained
with eight major oscillator shells, since convergence patterns with the number
of major oscillator shells that characterize the N$^3$LO, CD-Bonn, and $V_{18}$
interactions are
practically identical.

As expected, the CD-Bonn interaction gives more attraction than
N$3$LO, while the Argonne $V_{18}$ interaction
model yields less attraction than the other two models. The CD-Bonn potential has the weakest
tensor force of the three interactions studied here, whereas the $V_{18}$ interaction has the strongest
tensor force component. It is well-known that an interaction model with a weak
tensor force yields less quenching in the medium for the important  
$^3S_1$ and $^3D_1$ partial wave contributions to various matrix elements of the Hamiltonian.
The quenching is ascribed to both a Pauli effect and an energy dependence reflected in second- and 
higher-order terms (see, for example,
Ref.~\cite{hko95} for a discussion of this topic in both nuclei and nuclear matter). 
Although all interaction models fit properties of the deuteron and
the scattering data with a $\chi^2$ per datum close to $1$, the non-localities
which are introduced due to the way the interactions are constructed are responsible
for different results in a many-body context. 
\begin{table}[hbtp]
\caption{A comparison of the binding energies per particle
for $^{15}$O and $^{15}$N (the PR-EOMCCSD
values), $^{16}$O (the CCSD values), and $^{17}$O and
$^{17}$F (the PA-EOMCCSD values), obtained with the
N$^3$LO \cite{entem2003}, CD-Bonn \cite{cdbonn2000}, and
$V_{18}$ \cite{v18} potentials, and eight major oscillator shells,
with the experimental data taken from Ref.~\cite{audi2003}.
All entries are in MeV. For the CD-Bonn and N$^3$LO interactions,
we used $\hbar\Omega=11$ MeV. For $V_{18}$, we used
$\hbar\Omega=10$ MeV. For eight major shells, the results are practically
independent of the choice of $\hbar\Omega$ and $\beta_{\rm CoM} = 0.0$.
\label{tab:becomp}}
\begin{center}
\begin{tabular}{lcccc}
\hline
 & \multicolumn{3}{c}{Interaction} & \\
\cline{2-4}
Nucleus & N$^3$LO& CD-Bonn & $V_{18}$ & Expt \\
\hline
$^{15}$O& 6.158 & 6.643 & 4.789 & 7.464 \\ 
$^{15}$N& 6.339 & 6.810 & 4.957 & 7.699 \\
$^{16}$O& 6.951 & 7.444 & 5.469 & 7.976 \\
$^{17}$O& 6.722 & 7.201 & 5.214 & 7.751 \\
$^{17}$F& 6.559 & 7.048 & 5.059 & 7.542 \\
\hline
\end{tabular}
\end{center}
\end{table}
Indeed, the N$^3$LO and CD-Bonn models are non-local
interactions defined in momentum space.
While the N$^3$LO model is based on chiral Lagrangians with nucleons and pions
as degrees of freedom, including the
non-iterative $2\pi$ diagrams at chiral fourth order, the CD-Bonn interaction is
a traditional meson-exchange model that includes the six
low-mass mesons $\pi,\delta,\rho,\Omega,\eta$ and the
fictitious $\sigma$ meson, which is a $2\pi$ resonance. The
Argonne $V_{18}$ model is based on a local $r$-space parametrization, dominated by one-pion exchange.
The strength of the nuclear tensor force is intimately connected with the non-localities 
of the different nucleon-nucleon forces. Depending on how it is quenched in a many-body context, one may
get less or more attraction. The attractive part of, for example, the 
$^3S_1$ partial wave contribution is more attractive in the medium for an interaction with a weak
tensor than for one with a strong tensor force. Such features are clearly seen in
the coupled-cluster results reported in Table \ref{tab:becomp},
where the potential with the weakest tensor force, CD-Bonn, 
yields more binding than the two other models.

As mentioned previously, 
our results are practically converged as functions of the number of harmonic oscillator shells.
Based on our earlier work \cite{wloch05}, the triply excited
clusters and the related $3p\mbox{-}2h$ and $3h\mbox{-}2p$ excitations
in the particle-attaching and particle-removing $R_{\mu}^{(A + 1)}$ and $R_{\mu}^{(A - 1)}$
operators of the PA-EOMCC and PR-EOMCC theories are expected to have very
little impact on the calculated binding energies.
We claim therefore that, except for a small correction due to triples and a weak starting energy dependence 
\cite{wloch05}, the lack of agreement between coupled-cluster and experimental
binding energies is primarily due to the missing physics in our Hamiltonians.
The main conclusion that one can derive from the results of our
coupled-cluster calculations with different interactions is that
every nucleon-nucleon interaction model needs its own three-body potential. 
The Argonne group has derived sophisticated three-body interaction terms (see, for example, the extensive 
elaboration of Ref.~\cite{steve2001}). The parameters entering their three-body
interaction models are fitted to reproduce properties of light nuclei.
These three-body terms follow much of the same pion-exchange picture adopted in the construction
of the Argonne $V_{18}$ interaction.
For the CD-Bonn interaction one would need to derive three-body terms based on  
a meson-exchange picture, as outlined, for example, by the Bochum group \cite{gari1996}. However, no 
such model, which accompanies
this interaction, has been fully developed. The situation for models based on effective field theory is 
much better as three-body terms arise quite naturally at given orders in the expansion parameter
\cite{epelbaum2003}. Our coupled-cluster
results indicate that every interaction, due to different non-localities, has its own three-body
component reflected in different binding energies and different spin-orbit splittings (the
$(3/2)_1^{-} - (1/2)_1^-$ spacings in $^{15}$O and $^{15}$N and the
$(3/2)^+_1 - (5/2)^+_1$ spacings in $^{17}$O and $^{17}$F), as demonstrated in
Table \ref{tab:becomp}, which lists binding energies per nucleon,
and Table \ref{tab:spectraforces}, which lists the corresponding low-lying excited states
of the valence systems around $^{16}$O examined in this work.
\begin{table}[hbtp]
\caption{A comparison of the energies of
the low-lying excited states of $^{15}$O, $^{15}$N, $^{17}$O and
$^{17}$F, relative to the corresponding ground-state energies
(the $(1/2)_1^-$ states of $^{15}$O and $^{15}$N and
the $(5/2)^+_1$ states of $^{17}$O and $^{17}$F),
obtained with the PR-EOMCCSD ($^{15}$O and $^{15}$N)
and PA-EOMCCSD ($^{17}$O and $^{17}$F)
methods, the N$^3$LO \cite{entem2003}, CD-Bonn \cite{cdbonn2000}, and
$V_{18}$ \cite{v18} potentials, and eight major oscillator shells,
with the experimental data taken from Ref.~\cite{fire}.
All entries are in MeV. For the CD-Bonn and N$^3$LO interactions,
we used $\hbar\Omega=11$ MeV. For $V_{18}$, we used
$\hbar\Omega=10$ MeV. For eight major shells, the results are practically
independent of the choice of $\hbar\Omega$ and $\beta_{\rm CoM} = 0.0$.
\label{tab:spectraforces}}
\begin{center}
\begin{tabular}{lcccc}
\hline
& \multicolumn{3}{c}{Interaction} & \\
\cline{2-4}
Excited state & N$^3$LO  &CD-Bonn& $V_{18}$ & Expt \\
\hline
$^{15}$O $(3/2)_1^-$&  6.264 & 7.351 &  4.452 & 6.176 \\
$^{15}$N $(3/2)_1^-$&  6.318 & 7.443 &  4.499 & 6.323 \\
$^{17}$O $(3/2)^+_1$&  5.675 & 6.406 &  3.946 & 5.084 \\
$^{17}$O $(1/2)^+_1$& -0.025 & 0.311 & -0.390 & 0.870 \\
$^{17}$F $(3/2)^+_1$&  5.891 & 6.677 &  4.163 & 5.000 \\
$^{17}$F $(1/2)^+_1$&  0.428 & 0.805 &  0.062 & 0.495\\
\hline
\end{tabular}
\end{center}
\end{table}

As shown in Table \ref{tab:spectraforces}, the
CD-Bonn and the N$^3$LO models result in the largest spin-orbit splittings (much larger than
in the case of $V_{18}$). In order to examine this behavior in some detail,
we have computed all diagrams through third order in the $G$ matrix for
$\hbar\Omega=14$ MeV, using many-body perturbation theory as described in Ref.~\cite{hko95},
including folded diagrams to infinite order. For example,
at the Hartree-Fock level, which corresponds to the first order in the $G$ matrix,
the spin-orbit splittings for neutrons between
the two hole states in the $0p$ shell are
$4.85$ MeV, $4.41$ MeV, and $3.91$ MeV for the CD-Bonn, N$^3$LO and $V_{18}$
interaction models, respectively.
Since we are dealing with spin-isospin saturated systems, the results for protons
are almost the same. The Hartree-Fock term yields the largest contribution and receives important
contributions from the short-range two-body spin-orbit force. However, there is also
a considerable contribution to the splitting that originates from the
second-order $2h\mbox{-}1p$ and $2p\mbox{-}1h$ terms.
The corresponding second-order contributions are 1.81 MeV, 1.73 MeV and 1.35 MeV
for the same three interactions, respectively.
These perturbation theory estimates agree with the ways the
$(3/2)_1^{-} - (1/2)_1^-$ spacings in $^{15}$O and $^{15}$N and the
$(3/2)^+_1 - (5/2)^+_1$ spacings in $^{17}$O and $^{17}$F, obtained in
the corresponding PR-EOMCCSD and PA-EOMCCSD calculations, vary with
the interaction.
This analysis illustrates, at least to some extent, the role played by      
the quenching of the
tensor force via the second and higher-order terms in many-body perturbation theory
in different interaction models.
The perturbative results do not stabilize, however, as functions of the oscillator energy,
a result which is in close agreement with the findings reported by
Fujii {\em et al.}  \cite{fujii2004}. With increasing $\hbar\Omega$,
the single-particle splittings increase if
one uses an unperturbed harmonic oscillator basis. We defer thus from a
more elaborate analysis of many-body perturbation theory,
since it yields results of a rather limited interest.
The problems with many-body perturbation theory, such as the lack of a proper indication of 
convergence in terms
of $G$ and the difficulties with going beyond third order in the interaction, are well known.
The coupled-cluster methods, including the quantum chemistry inspired CCSD, PA-EOMCCSD, and
PR-EOMCCSD approximations used in this work, are capable of summing large classes of diagrams to infinite order,
eliminating many of the problems encountered in many-body perturbation theory calculations,
and providing a much more stable description of the ground and excited states
of the valence systems around ${^{16}}$O. The differences between
coupled-cluster results obtained with different interactions point to the need for
developing three-body interactions consistent with a given two-body interaction model.

It is interesting to note that
in spite of the apparent differences between the converged coupled-cluster results obtained with
different pairwise interaction models, the relative binding energies
of $^{15}$O, $^{15}$N, $^{16}$O, $^{17}$O, and $^{17}$F
obtained with different interactions are in good agreement with experiment and with each other.
For example, as already mentioned the difference between experimental binding energies of
$^{16}$O and $^{17}$O is $0.225$ MeV per particle.
The CCSD and PA-EOMCCSD ground-state energies of $^{16}$O and $^{17}$O
resulting from the calculations with eight major oscillator shells
differ by 0.229 MeV per particle for N$^3$LO, 0.243 MeV per particle for CD-Bonn,
and 0.255 MeV per particle for $V_{18}$. Similarly, the
difference between experimental binding energies of $^{16}$O and $^{15}$O is
0.512 MeV per particle, whereas the CCSD and PR-EOMCCSD ground-state energies of
$^{16}$O and $^{15}$O differ by 0.793, 0.801, and 0.680 MeV per particle
for the N$^3$LO, CD-Bonn, and $V_{18}$ potentials, respectively. Here, the differences
with experiment are somewhat greater than in the case of $^{16}$O and $^{17}$O,
but the overall agreement among different potentials is still very good.
The differences between the binding energies for the $A=15$ nuclei and for the $A=17$ nuclei
obtained with different interactions are close to one another and to the experimental
values too. According to Table \ref{tab:becomp}, the experimental value of the
binding energy difference ${\rm BE(^{15}N) - BE(^{15}O)}$ is 0.235 MeV per particle.
The PR-EOMCCSD calculations with the N$^3$LO, CD-Bonn, and $V_{18}$ interactions give
0.181, 0.167, and 0.168 MeV per particle, respectively, for the same
binding energy difference. Similarly, the experimental value of the
binding energy difference ${\rm BE(^{17}O) - BE(^{17}F)}$ is 0.209 MeV per particle.
The PA-EOMCCSD calculations with the N$^3$LO, CD-Bonn, and $V_{18}$ potentials give
0.163, 0.153, and 0.155 MeV per particle, respectively, for
the same binding energy difference. In spite of the substantial
differences between binding energies resulting from the calculations
with different interactions, which are affected by the three-body forces that
are expected to be different for different pairwise interactions,
the binding energies per nucleon resulting from our
PR-EOMCCSD/CCSD/PA-EOMCCSD calculations with eight major oscillator shells
satisfy $^{15}{\rm O} < \: ^{15}{\rm N} < \: ^{17}{\rm F} < \: ^{17}{\rm O} < \: ^{16}{\rm O}$,
independent of the interaction used in coupled-cluster calculations.
This means that once we adjust the value of the binding energy of the reference
$^{16}{\rm O}$ system, we can obtain the interaction-independent ordering
of the binding energies
of the valence nuclei around $^{16}{\rm O}$.
With an exception of the $^{15}{\rm N}$ and $^{17}{\rm F}$ nuclei, whose
binding energy ordering should be reversed, the ordering of binding energies
per particle resulting from the
relatively inexpensive coupled-cluster calculations is in good agreement
with experiment.
These are encouraging findings from the point of view of the future applications
of coupled-cluster methods employing renormalized Hamiltonians in nuclear physics.

\section{Conclusions and Perspectives}
\label{sec:conclusions}

We summarize here our main conclusions and perspectives for future studies.

\begin{enumerate}
\item To our knowledge, this is the first application of the {\em ab initio} coupled-cluster theory
employing the renormalized form of the Hamiltonian,
combined with the PA-EOMCC and PR-EOMCC formalisms for open-shell many-fermion systems,
to nuclear valence systems with one valence particle or
one valence hole. We have shown that one can obtain virtually converged
results with given two-body Hamiltonians for both binding energies and
low-lying excited states. This has been possible thanks to the development
of highly efficient CCSD, PA-EOMCCSD, and PR-EOMCCSD computer codes
and the use of the renormalized Hamiltonians in our calculations, which
lead to a rapid convergence with the number of oscillator shells in a basis.
The systems whose properties have been studied in this work were
$^{15}$O, $^{15}$N, $^{17}$O and
$^{17}$F. An emphasis has been placed on states
dominated by one-quasi-particle configurations.
The discrepancies between the results of large-scale coupled-cluster calculations
for these nuclei and the corresponding experimental data have been traced to the Hamiltonians
used in the calculations, much less to the correlations neglected in coupled-cluster
approximations employed in this study.
\item Three different nucleon-nucleon interactions have been used to define our two-body Hamiltonians.
These are the N$^3$LO model
\cite{entem2003}, the CD-Bonn interaction 
\cite{cdbonn2000}, and
the $V_{18}$ model of the Argonne group \cite{v18}.
All of these interactions yield different binding energies and different energies of the 
excited states. The different 
binding energies and spin-orbit splittings can
be related to varying non-localities in the nucleon-nucleon interactions.
Of particular interest here has been the role played by the nuclear tensor force.
The different behavior of the three interaction models examined in this study points
to the need for the development of the interaction specific three-body forces.
\item We have also demonstrated that most of the discrepancy between theory and experiment
for the $1p\mbox{-}1h$ negative parity states 
in $^{16}$O, including the lowest $3^{-}_{1}$ state examined in our earlier work \cite{wloch05},
can be retraced to the difference between the theoretical and experimental values
of the relevant energy gaps between neutron or proton 
states in the $0p$ and $1s0d$ shells. 
\item In spite of the differences among interactions, the relative binding energies
of the $^{15}{\rm O}$, $^{15}{\rm N}$, $^{17}{\rm F}$, $^{17}{\rm O}$, and $^{16}{\rm O}$
resulting from the coupled-cluster calculations seem to be virtually independent of
the interaction and in good agreement with experiment. The
$(3/2)_1^{-} - (1/2)_1^-$ spacings in $^{15}$O and $^{15}$N resulting from the
converged coupled-cluster calculations with the N$^3$LO interaction are in 
excellent agreement with experiment, indicating that
the spin-orbit force associated with an
eventual three-body force for N$^3$LO should be small.
\end{enumerate}

There are several obvious extensions to this work. First of all,
the need for an inclusion of three-body interactions sets the agenda for
forthcoming studies. Moreover, it may be useful to examine
the role of $3p\mbox{-}2h$ and $3h\mbox{-}2p$ correlations
in the PA-EOMCC and PR-EOMCC calculations, which we neglected in this study.
For the states considered here, the $3p\mbox{-}2h$ and $3h\mbox{-}2p$ correlations
are expected to be small, since the states of $^{15}{\rm O}$, $^{15}{\rm N}$, $^{17}{\rm F}$,
and $^{17}{\rm O}$ that we have examined show relatively
small departures from an independent-particle picture and since the underlying $T_{3}$ cluster
contributions that define the reference $^{16}{\rm O}$ system are small \cite{wloch05}. On the other hand,
we have tacitly assumed that the $0d_{3/2}$ states of 
$^{17}$F and $^{17}$O are bound states. These states are resonances,
and it is not yet entirely clear how the non-resonant continuum may affect
the description of these states. The inclusion of such contributions in the description of these states 
is another important point to explore, as demonstrated
in the recent works on the Gamow shell-model and complex-scaling techniques \cite{witek,roberto,gaute}.

\section*{Acknowledgements}

This work has been supported by the Chemical Sciences, Geosciences
and Biosciences Division, Office of Basic Energy Sciences, Office
of Science, U.S. Department of Energy (Grant No. DE-FG02-01ER15228) and
the National Science Foundation (Grant No. CHE-0309517;
both awards provided to P.P.). Additional support in the form
of the Dissertation Completion Fellowship provided by
Michigan State University to J.R.G is acknowledged too.
J.R.G. is a National Science Foundation Graduate Research Fellow.
This research is also supported by the Office of Nuclear Physics, 
Office of Science of the U.S. Department of Energy  under 
Contract Number DE-AC05-00OR22725 with UT-Battelle, LLC (Oak 
Ridge National Laboratory).  This research is also supported by 
the Research Council of Norway (Program
for Supercomputing; a grant of computing time).

The calculations reported here were performed on the 
computer systems provided by the High Performance Computing Center 
at Michigan State University, the National Energy Research Scientific 
Computing Center at Laurence Berkeley National Laboratory, and the 
Center for Computational Science at Oak Ridge National Laboratory. 

\appendix*

\section{Factorized form of the CCSD, PA-EOMCCSD, and PR-EOMCCSD
equations}

In this appendix, we present the working equations defining the
CCSD, PA-EOMCCSD, and PR-EOMCCSD methods
exploited in this study. All of the equations are
expressed in terms of the one- and two-body
matrix elements of the Hamiltonian in the normal-ordered form,
$f_{\alpha}^{\beta}$ and $v_{\alpha\beta}^{\gamma\delta}$, respectively (cf. Eq. (\ref{hnormal});
in our case, $f_{\alpha}^{\beta}$ and $v_{\alpha\beta}^{\gamma\delta}$ are the
one- and two-body matrix elements of the normal-ordered form of the effective Hamiltonian
$H$, Eq. (\ref{hfinal})),
the $t_{a}^{i}$ and $t_{ab}^{ij}$ cluster amplitudes defining the underlying
$A$-particle ground-state CCSD problem, and, in the
case of the PA-EOMCCSD and PR-EOMCCSD approaches,
the $r_{a}$, $r_{ab}^{\hspace{4 pt} j}$, $r^{i}$, and $r^{ij}_{\hspace{3 pt} b}$
amplitudes defining the particle-attaching ($r_{a}$ and $r_{ab}^{\hspace{4 pt} j}$)
and particle-removing ($r^{i}$ and $r^{ij}_{\hspace{3 pt} b}$) operators,
$R_{\mu}^{(A+1)}(2p\mbox{-}1h)$ and $R_{\mu}^{(A-1)}(2h\mbox{-}1p)$,
Eqs. (\ref{r2p1h}) and (\ref{r2h1p}), respectively.
As explained in Sec.~\ref{subsec:eomccsd}, the CCSD, PA-EOMCCSD, and PR-EOMCCSD
equations can be cast into a computationally efficient
factorized form expressed in terms of the one- and two-body
matrix elements of the CCSD similarity-transformed Hamiltonian
$\bar{H}_{N,\rm{open}}{\rm{(CCSD)}}$,
$\bar{h}_{\alpha}^{\beta}$ and $\bar{h}_{\alpha\beta}^{\gamma\delta}$,
respectively, and a few additional
intermediates that are generated in a recursive manner.
The complete set of one- and two-body matrix elements
of $\bar{H}_{N,\rm{open}}({\rm CCSD})$ and
other intermediates that are needed to set up the
CCSD, PA-EOMCCSD, and PR-EOMCCSD equations is given
in Table \ref{tableappendix}.
\begin{table}
\caption{
\label{tableappendix}
Explicit algebraic expressions for the one- and two-body
matrix elements elements of $\bar{H}_{N,\rm{open}}({\rm CCSD})$
($\bar{h}_{\alpha}^{\beta}$ and $\bar{h}_{\alpha\beta}^{\gamma\delta}$,
respectively)
and other intermediates (designated by $I$) used
to construct the computationally efficient form
of the CCSD, PA-EOMCCSD, and PR-EOMCCSD equations.
}
\begin{tabular}{l @{\extracolsep{0.1in}} l}
\hline
Intermediate & Expression\footnotemark[1] \\
\hline
$\bar{h}_{i}^{a}$ & $f_{i}^{a} + v_{im}^{ae} t_{e}^{m} $ \\
$\bar{h}_{i}^{j}$ & $f_{i}^{j} + v_{im}^{je} t_{e}^{m}
+ \half v_{mi}^{ef} t_{ef}^{mj} + \bar{h}_{i}^{e} t_{e}^{j}$ \\
$\bar{h}_{a}^{b}$ & $I_{a}^{b} - \bar{h}_{m}^{b} t_{a}^{m}$ \\
$\bar{h}_{ai}^{bc}$ & $v_{ai}^{bc} - v_{mi}^{bc} t_{a}^{m}$ \\
$\bar{h}_{ij}^{ka}$ & $v_{ij}^{ka} + v_{ij}^{ea} t_{e}^{k}$ \\
$\bar{h}_{ab}^{cd}$ & $v_{ab}^{cd} + \half v_{mn}^{cd} t_{ab}^{mn}
- \bar{h}_{am}^{cd} t_{b}^{m} + v_{bm}^{cd} t_{a}^{m}$ \\
$\bar{h}_{ij}^{kl}$ & $v_{ij}^{kl} + \half
v_{ij}^{ef} t_{ef}^{kl} - \bar{h}_{ij}^{le} t_{e}^{k} + v_{ij}^{ke} t_{e}^{l}$ \\
$\bar{h}_{ia}^{jb}$ & $I_{ia}^{\p jb} - v_{im}^{eb} t_{ea}^{jm}
- \bar{h}_{im}^{jb} t_{a}^{m}$ \\
$\bar{h}_{ab}^{ic}$ & $v_{ab}^{ic} + v_{ab}^{ec} t_{e}^{i}
- \bar{h}_{mb}^{ic} t_{a}^{m} + I_{ma}^{\p ic} t_{b}^{m}$
\\
& $ - \bar{h}_{m}^{c} t_{ab}^{im}
+ \bar{h}_{bm}^{ce} t_{ae}^{im} - v_{am}^{ce} t_{be}^{im} + \half \bar{h}_{nm}^{ic} t_{ab}^{nm}$
\\
$\bar{h}_{ia}^{jk}$ & $v_{ia}^{jk} + \bar{h}_{mi}^{jk} t_{a}^{m}
- v_{ia}^{ke} t_{e}^{j}
+ {\mathscr A}^{jk} \bar{h}_{im}^{je} t_{ae}^{km}$
\\
& $ + \bar h_{i}^{e} t_{ea}^{jk} + I_{ia}^{\p je} t_{e}^{k}
- \half v_{ai}^{ef} t_{ef}^{jk}$
\\
$I_{a}^{\p b}$ & $f_{a}^{b} + v_{am}^{be} t_{e}^{m}$ \\
$I_{a}^{b}$ & $I_{a}^{\p b} - \half v_{mn}^{eb} t_{ea}^{mn}$ \\
$I_{ab}^{ic}$ & $v_{ab}^{ic} + v_{ab}^{ec} t_{e}^{i} + v_{mb}^{ec} t_{ae}^{im}$ \\
$I_{ia}^{jk}$ & $\bar{h}_{ia}^{jk} - \half \bar{h}_{mi}^{jk} t_{a}^{m}$ \\
$I_{ia}^{jb}$ & $-\half v_{im}^{eb} t_{ea}^{jm}$ \\
$I_{ia}^{\p jb}$ & $v_{ia}^{jb} + v_{ia}^{eb} t_{e}^{j}$ \\
$I_{m}$ & $\half v_{mn}^{ef}r_{ef}^{\hspace{4pt}n}$ \\
$I^{e}$ & $-\half v_{mn}^{ef}r^{mn}_{\hspace{5pt}f}$ \\
\hline
\end{tabular}
\footnotetext[1]{
Summation over repeated upper and lower indices is assumed.
$f_{\alpha}^{\beta}=\langle \alpha |f| \beta \rangle$ and
$v_{\alpha\beta}^{\gamma\delta}
= \langle \alpha\beta |v| \gamma\delta \rangle
- \langle \alpha\beta |v| \delta\gamma \rangle $
are the one- and two-body matrix elements of the
Hamiltonian in the normal-ordered form, Eq. (\ref{hnormal}) and
the $t_{a}^{i}$ and $t_{ab}^{ij}$ are the
singly and doubly excited cluster amplitudes defining the
ground-state CCSD wave function of the $A$-body reference system.
}
\end{table}

The ground-state CCSD equations for the singly and doubly excited
cluster amplitudes $t_{a}^{i}$ and $t_{ab}^{ij}$, Eqs. (\ref{momccsd1})
and (\ref{momccsd2}), can be given the following, computationally
efficient form:
\begin{eqnarray}
\bar{h}_{a}^{i} & \equiv & f_a^i + I^{\p e}_a t_e^i - \bar{h}_m^i t_a^m - v_{ma}^{ie} t_{e}^{m}
+ \bar{h}_m^e t_{ea}^{mi} - \half \bar{h}_{mn}^{ie} t_{ae}^{mn}
\nonumber
\\
& &
+ \half v_{am}^{ef} t_{ef}^{im} = 0 ,
\label{eqccsd1}
\end{eqnarray}
\begin{eqnarray}
\bar{h}_{ab}^{ij} & \equiv & v_{ab}^{ij} + {\mathscr A}_{ab} {\mathscr A}^{ij}
[ \half I^{ie}_{ab} t_{e}^{j}
- \half I_{mb}^{ij} t_{a}^{m} + \half I^{e}_{b} t_{ae}^{ij}
\nonumber
\\
& &
+ \eight v_{ab}^{ef} t_{ef}^{ij}
+ \eight \bar{h}_{mn}^{ij} t_{ab}^{mn}
- I^{ie}_{mb} t_{ae}^{mj}
- \half \bar{h}_{m}^{j} t_{ab}^{im} ]
\nonumber
\\
& = & 0 .
\label{eqccsd2}
\end{eqnarray}
Note that the left-hand sides of Eqs. (\ref{momccsd1})
and (\ref{momccsd2}) (or (\ref{eqccsd1}) and (\ref{eqccsd2})) represent,
respectively, the one- and two-body matrix elements
$\bar{h}_{a}^{i}$ and $\bar{h}_{ab}^{ij}$ of $\bar{H}_{N}({\rm CCSD})$
(see Eqs. (\ref{ccsd1}) and (\ref{ccsd2})).
The relevant intermediates can be found in Table \ref{tableappendix}.
The antisymmetrizers ${\mathscr A}_{pq}={\mathscr A}^{pq}$,
which enter Eq. (\ref{eqccsd2}) and other equations presented
in this appendix, are defined as
\beq
{\mathscr A}_{pq} \equiv {\mathscr A}^{pq} = 1 - (pq),
\eeq
with $(pq)$ representing a transposition of two
indices.
Once the above equations are solved for $t_{a}^{i}$ and $t_{ab}^{ij}$,
the ground-state CCSD energy is calculated using the formula
(cf. Eq. (\ref{egratwobody}))
\beq
E_{0}^{(A)}(M) = \langle \Phi | H | \Phi \rangle +
f_{i}^{a} t_{a}^{i} + \qua v_{ij}^{ab} (t_{ab}^{ij} + 2 t_{a}^{i} t_{b}^{j}),
\label{energyccsd}
\eeq
which is valid for any truncation scheme $M \geq 2$.

Once the $t_{a}^{i}$ and $t_{ab}^{ij}$ amplitudes are determined and
the ground-state CCSD energy of the reference $A$-body system is known,
we can set up and solve the eigenvalue equations defining the PA-EOMCCSD and
PR-EOMCCSD methods. The PA-EOMCCSD equations for the energy differences
$\omega_{\mu}^{(A + 1)} = E_{\mu}^{(A+1)} - E_{0}^{(A)}$ and
the $1p$ and $2p\mbox{-}1h$ amplitudes, $r_{a}$ and $r_{ab}^{\hspace{4 pt} j}$,
respectively, defining the ground and excited states of the
$(A+1)$-particle system, can be given the following, computationally efficient,
form:
\begin{widetext}
\beq
\langle \Phi^{a} | [\bar{H}_{N,\rm{open}}{\rm{(CCSD)}} R_{\mu}^{(A+1)}(2p\mbox{-}1h)]_{C} |\Phi\rangle
= \bar{h}_{a}^{e}r_{e}
+\bar{h}_{m}^{e}r_{ae}^{\hspace{3 pt}m}
+\half \bar{h}_{am}^{ef}r_{ef}^{\hspace{3 pt}m}
= \omega_{\mu}^{(A+1)} \, r_{a} ,
\label{work1}
\eeq
\begin{eqnarray}
\langle \Phi_{\hspace{4 pt}j}^{ab} | [\bar{H}_{N,\rm{open}}{\rm{(CCSD)}}
R_{\mu}^{(A+1)}(2p\mbox{-}1h)]_{C} |\Phi\rangle
& = & {\mathscr A}_{ab}\lbrack -\half\bar{h}_{ab}^{je}r_{e}+
\bar{h}_{a}^{e}r_{eb}^{\hspace{3 pt}j}
-\half\bar{h}_{m}^{j}r_{ab}^{\hspace{3 pt}m}
+\qua \bar{h}_{ab}^{ef}r_{ef}^{\hspace{4 pt}j}
-\bar{h}_{ma}^{je}r_{eb}^{\hspace{3 pt}m}
-\half I_{m}t_{ab}^{mj} \rbrack
\nonumber
\\
& = &
\omega_{\mu}^{(A+1)} \, r_{ab}^{\hspace{4 pt} j} .
\label{work2}
\end{eqnarray}
\end{widetext}
Similarly, we can use the CCSD values of the
singly and doubly excited cluster amplitudes
defining the ground-state wave function of the reference $A$-body system
to set up the PR-EOMCCSD eigenvalue equations for the
energy differences $\omega_{\mu}^{(A - 1)} = E_{\mu}^{(A-1)} - E_{0}^{(A)}$
and the $1h$ and $2h\mbox{-}1p$ amplitudes,
$r^{i}$ and $r^{ij}_{\hspace{3 pt} b}$, respectively, defining
the ground and excited states of the $(A-1)$-particle system.
The computationally efficient form of the
PR-EOMCCSD equations is as follows:
\begin{widetext}
\beq
\langle \Phi_{i} | [ \bar{H}_{N,\rm{open}}{\rm{(CCSD)}} R_{\mu}^{(A-1)}(2h\mbox{-}1p)]_{C} |\Phi\rangle
= -\bar{h}_{m}^{i}r^{m}
+\bar{h}_{m}^{e}r^{im}_{\hspace{5 pt}e}
-\half \bar{h}_{mn}^{ie}r^{mn}_{\hspace{5pt}e}
=
\omega_{\mu}^{(A-1)} \, r^{i} ,
\label{work4}
\eeq
\begin{eqnarray}
\langle \Phi^{\hspace{4 pt}b}_{ij} | [ \bar{H}_{N,\rm{open}}{\rm{(CCSD)}}
R_{\mu}^{(A-1)}(2h\mbox{-}1p)]_{C} |\Phi\rangle
& = &
{\mathscr A}^{ij}\lbrack -\half\bar{h}_{mb}^{ij}r^{m}-
\bar{h}_{m}^{i}r^{mj}_{\hspace{5pt}b}
+\half\bar{h}_{b}^{e}r^{ij}_{\hspace{4pt}e}
+\qua \bar{h}_{mn}^{ij}r^{mn}_{\hspace{5pt}b}
-\bar{h}_{mb}^{ie}r^{mj}_{\hspace{5pt}e}
+\half I^{e}t_{eb}^{ij} \rbrack
\nonumber
\\
& = &
\omega_{\mu}^{(A-1)} \, r^{ij}_{\hspace{3 pt} b} ,
\label{work5}
\end{eqnarray}
\end{widetext}
%


\begin{thebibliography}{999}

\bibitem{pp1993}
S.C.~Pieper and V.R.~Pandharipande, Phys.~Rev.~Lett.~{\bf 70}, 
2541 (1993).

\bibitem{ab1981}
K.~And$\rm \overline{o}$ and H.~Band$\rm \overline{o}$, Prog.~Theor.~Phys.~{\bf 66}, 
227 (1981).

\bibitem{pieper02}
R.B.~Wiringa and S.C.~Pieper, Phys. Rev. Lett. {\bf 89}, 182501 (2002).

\bibitem{navratil02}
P.~Navr{\'a}til and W.E. Ormand, Phys. Rev. C {\bf 68}, 034305 (2003);
Phys. Rev. Lett.~{\bf 88}, 152502 (2002).

\bibitem{hko95}
M.~Hjorth-Jensen, T.T.S.~Kuo, and E.~Osnes,
Phys.\ Rep.\ {\bf 261}, 125 (1995).

\bibitem{brown2001}
B.A.~Brown, 
Prog.~Part.~Nucl.~Part.~{\bf 47}, 517  (2001), and references therein.

\bibitem{otsuka2004}
M.~Homna, T.~Otsuka, B.A.~Brown, and T.~Mizusaki,
Phys. Rev. C {\bf 69}, 034335 (2004).

\bibitem{brown2002}
B.A.~Brown, 
Prog.~Theor.~Phys.~Supp.~{\bf 146}, 23 (2002).

\bibitem{coester}
F. Coester, Nucl. Phys. {\bf 7}, 421 (1958).

\bibitem{coesterkummel}
F. Coester and H. K{\" u}mmel, Nucl. Phys. {\bf 17},
477 (1960).

\bibitem{cizek}
J.~{{\v C}{\'\i}{\v z}ek}, J. Chem. Phys. {\bf 45}, 4256 (1966);
Adv. Chem. Phys. {\bf 14}, 35 (1969).

\bibitem{palduscizek}
J.~{\v C}{\'\i}{\v z}ek and
J.~Paldus, Int. J. Quantum Chem. {\bf 5}, 359 (1971).

\bibitem{chem_rev1}
J.~Paldus and X.~Li, Adv. Chem. Phys. \textbf{110}, 1 (1999).

\bibitem{chem_rev2}
J.~Paldus, in {\em Theory and Applications of Computational Chemistry: The
First 40 Years}, edited by C.~F.~Dykstra, G.~Frenking, K.~S.~Kim, and
G.~E.~Scuseria (Elsevier, Amsterdam, 2005), in press.

\bibitem{chem_rev3}
R.J.~Bartlett and J.F.~Stanton, Rev. Comput. Chem. \textbf{5}, 65 (1994).

\bibitem{chem_rev4}
R.J.~Bartlett, in {\em Modern Electronic
Structure Theory}, Part I, edited by D.~R.~Yarkony (World Scientific, Singapore, 1995),
p. 1047.

\bibitem{chem_rev5}
T.D.~Crawford and H.F. Schaefer III, Rev. Comput. Chem. \textbf{14}, 33 (2000).

\bibitem{chem_rev6}
J.~Gauss, in {\em Encyclopedia of Computational Chemistry}, edited by
P.v.R.~Schleyer, N.L.~Allinger, T. Clark,
J.~Gasteiger, P.A.~Kollman, H.F.~Schaefer III, and P.R.~Schreiner
(Wiley, Chichester, 1998), Vol. 1, p. 615.

\bibitem{chem_rev7}
J.~Paldus, in {\em Handbook of Molecular Physics and Quantum Chemistry},
edited by S.~Wilson (Wiley, Chichester, 2003), Vol. 2, p. 272.

\bibitem{piecuch2002}
P.~Piecuch, K.~Kowalski, I.S.O.~Pimienta, and M.J.~ McGuire, Int.
Rev. Phys. Chem. {\bf 21}, 527 (2002).

\bibitem{piecuch2003}
P.~Piecuch, K.~Kowalski, P.-D.~Fan, and I.S.O.~Pimienta, in
{\em Advanced Topics in Theoretical Chemical Physics}, Vol. 12 of
{\em Progress in Theoretical Chemistry and Physics}, edited by
J.~Maruani, R.~Lefebvre , and E.~Br{\" a}ndas,
(Kluwer, Dordrecht, 2003), p. 119.

\bibitem{piecuch2004}
P.~Piecuch, K.~Kowalski, I.S.O.~Pimienta, P.-D.~Fan,
M.~Lodriguito, M.~J.~McGuire, S.~A.~Kucharski, T.~Ku{\' s}, and
M.~Musia{\l}, Theor. Chem. Acc. {\bf 112}, 349 (2004).

\bibitem{piecuch2005}
P.~Piecuch, M.~W{\l}och, M.~Lodriguito, and J.R.~Gour,
in {\em Progress in Theoretical Chemistry and Physics},
Vol. XXX, {\em Recent Advances in the Theory of Chemical and Physical Systems},
edited by
S.~Wilson, J.-P.~Julien, J.~Maruani, E.~Br{\" a}ndas, and G.~Delgado-Barrio,
(Springer, Berlin, 2005), p. XXX, in press.

\bibitem{pr1978}
H.~K{\" u}mmel,
K.H.~L{\" u}hrmann, and J.G.~Zabolitzky, Phys. Rep.
{\bf 36}, 1 (1978).

\bibitem{kummel2002}
H.G. K\"ummel,
in \textit{Recent Progress in Many-Body Theories},
edited by R.F. Bishop {\it et al.} (World Scientific, Singapore, 2002),
p. 334.

\bibitem{aggravatedrectifier97}
R.F~. Bishop, in \textit{Microscopic Quantum Many-Body
Theories and Their Applications}, edited by J. Navarro and A.
Polls, Lecture Notes in Physics, Vol. 510 (Springer, Berlin, 1998), p. 1.

\bibitem{v18}
R.B.~Wiringa, V.G.J.~Stoks, and R.~Schiavilla, Phys. Rev. C {\bf 51},
38 (1995).

\bibitem{cdbonn2000}
R.~Machleidt, Phys. Rev. C {\bf 63}, 024001 (2001).

\bibitem{entem2002}
D.R.~Entem and R.~Machleidt, Phys. Lett. B {\bf 524}, 93 (2002).

\bibitem{entem2003}
D.R.~Entem and R.~Machleidt, Phys. Rev. C {\bf 68}, 41001 (2003).

\bibitem{bogdan1}
B.~Mihaila and J.H.~Heisenberg, Phys. Rev. Lett.~{\bf 84}, 1403 (2000).

\bibitem{bogdan2}
B.~Mihaila and J.H.~Heisenberg, Phys.~Rev.~C {\bf 61}, 054309 (2000).

\bibitem{bogdan3}
B.~Mihaila and J.H.~Heisenberg, Phys. Rev. C {\bf 60}, 054303 (1999).

\bibitem{bogdan4}
J.H.~Heisenberg, and B.~Mihaila, Phys. Rev.~C {\bf 59}, 1440 (1999).

\bibitem{dean04}
D.J.~Dean and M.~Hjorth-Jensen, Phys.~Rev.~C {\bf 69}, 054320 (2004).

\bibitem{kowalski04}
K.~Kowalski, D.J.~Dean, M.~Hjorth-Jensen, T.~Papenbrock, and P.~Piecuch,
Phys.~Rev.~Lett.~{\bf 92}, 132501 (2004).


\bibitem{wloch05}
M.~W{\l}och, D.J.~Dean, J.R.~Gour, M.~Hjorth-Jensen, K.~Kowalski, 
T.~Papenbrock, and P.~Piecuch,
Phys.~Rev.~Lett.~{\bf 94}, 212501 (2005).

\bibitem{dean05}
D.J. Dean, J.R. Gour, G. Hagen, M. Hjorth-Jensen, K. Kowalski, T. Papenbrock,
and P. Piecuch, Nucl. Phys. A {\bf 752}, 299 (2005).

\bibitem{epja2005}
M.~W{\l}och, D.J.~Dean, J.R.~Gour, P.~Piecuch, M.~Hjorth-Jensen,
T.~Papenbrock, and K.~Kowalski,
Eur. Phys. J. A Direct, published online
on May 13, 2005;
DOI: 10.1140/epjad/i2005-06-062-8.

\bibitem{jpg2005}
M.~W{\l}och, J.R.~Gour, P.~Piecuch, D.J.~Dean, M.~Hjorth-Jensen, and T.~Papenbrock,
J. Phys. G: Nucl. Part. Phys. {\bf 31}, S1291 (2005).

\bibitem{navratil2004}
P.~Navr{\'a}til and J.P.~Vary, private communication 2004.

\bibitem{sak1}
S.A.~Kucharski and R.J.~Bartlett, Theor. Chim. Acta {\bf 80}, 387 (1991).

\bibitem{ccgamess}
P.~Piecuch, S.A.~Kucharski, K.~Kowalski, and M.~Musia{\l},
Comp. Phys. Commun. {\bf 149}, 71 (2002).

\bibitem{wloch2005}
M.~ W{\l}och, J.R.~Gour, K.~Kowalski, and P.~Piecuch,
J. Chem. Phys. {\bf 122}, 214107 (2005).

\bibitem{oureom}
P.~Piecuch and R.J.~Bartlett, Adv. Quantum Chem.~{\bf 34}, 295 (1999).

\bibitem{eaccsd1}
M. Nooijen and R.J. Bartlett, J. Chem. Phys. {\bf 102}, 3629 (1995).

\bibitem{eaccsd2}
M. Nooijen and R.J. Bartlett, J. Chem. Phys. {\bf 102}, 6735 (1995).

\bibitem{eaccsdt1}
M. Musia{\l} and R.J. Bartlett, J. Chem. Phys. {\bf 119}, 1901 (2003).

\bibitem{jeffeaip}
J.R. Gour, P. Piecuch, and M. W{\l}och, J. Chem. Phys., submitted.

\bibitem{ipccsd2}
M. Nooijen and J.G. Snijders, Int. J. Quantum Chem. Symp. {\bf 26}, 55 (1992).

\bibitem{ipccsd3}
M. Nooijen and J.G. Snijders, Int. J. Quantum Chem. {\bf 48}, 15 (1993).

\bibitem{ipccsd4}
J.F. Stanton and J. Gauss, J. Chem. Phys. {\bf 101}, 8938 (1994).

\bibitem{ipccsdt1}
M. Musia{\l}, S.A. Kucharski, and R.J. Bartlett,
J. Chem. Phys. {\bf 118}, 1128 (2003).

\bibitem{ipccsdt2}
M. Musia{\l} and R.J. Bartlett, Chem. Phys. Lett. {\bf 384}, 210 (2004).

\bibitem{ipccsdt-3}
Y.J. Bomble, J.C. Saeh, J.F. Stanton, P.G. Szalay, M. K{\' a}llay, and
J. Gauss, J. Chem. Phys. {\bf 122}, 154107 (2005).

\bibitem{eomcc1}
J. Geertsen, M. Rittby, and R.J. Bartlett, Chem. Phys. Lett.
{\bf 164}, 57 (1989).

\bibitem{eomcc3}
J.F. Stanton and R.J. Bartlett, J. Chem. Phys.
{\bf 98}, 7029 (1993).

\bibitem{eft}
S. Weinberg, Phys.~Lett.~B {\bf 251}, 288 (1990).

\bibitem{bira1999}
U.~van Kolck, Prog.~Part.~Nucl.~Phys.~{\bf 43}, 337 (1999).

\bibitem{dean99}
D. J. Dean, M. T. Ressell, M. Hjorth-Jensen, 
S. E. Koonin, K. Langanke, and A. P. Zuker,
Phys. Rev. C {\bf 59}, 2474 (1999)

\bibitem{bbp63}
H.A.~Bethe, B.H.~Brandow, and A.G.~Petschek,
{\em Phys. Rev.}, {\bf 129}, 225 (1963).

\bibitem{leesuzuki1}
S.Y.~Lee and K. Suzuki, {\em Phys. Lett. B}, {\bf 91}, 79 (1980).

\bibitem{leesuzuki2}
K.~Suzuki and S.Y.~Lee, {\em Prog. Theor. Phys.}, {\bf 64}, 2091 (1980).

\bibitem{zuker1}
A. Poves, and A. Zuker, Phys. Rep. {\bf 71}, 141 (1981). 

\bibitem{vlowk}
S.K.~Bogner, T.T.S.~Kuo, and A.~Schwenk, Phys.~Rep.~{\bf 386},
1 (2003).

\bibitem{cct}
J.~Hubbard, Proc. Roy. Soc. (London) A {\bf 240}, 539 (1957); {\bf 243}, 336 (1958);
{\bf 244}, 199 (1959).

\bibitem{lct1}
K.A.~Brueckner, Phys.~Rev.~Lett.~{\bf 100}, 36 (1955).

\bibitem{lct2}
J.~Goldstone, Proc. Roy. Soc. (London) A {\bf 239}, 267 (1957).

\bibitem{lct3}
N.M.~Hugenholtz, Physics (Utrecht) {\bf 23}, 481 (1957).

\bibitem{purvis82}
G.D.~Purvis III and R.J.~Bartlett, J. Chem. Phys. {\bf 76}, 1910 (1982).

\bibitem{ccsdfritz}
G.E.~Scuseria, A.C.~Scheiner, T.J.~Lee, J.E.~Rice, and
H.F.~Schaefer III, J. Chem. Phys. {\bf 86}, 2881 (1987).

\bibitem{osaccsd}
P.~Piecuch and J.~Paldus, Int. J. Quantum Chem. {\bf 36},
429 (1989).

\bibitem{leszcz}
P.~Piecuch and K.~Kowalski, in {\em Computational Chemistry:
Reviews of Current Trends}, edited by J. Leszczy{\' n}ski (World
Scientific, Singapore, 2000), Vol. 5, p. 1.

\bibitem{pulay1}
P.~Pulay, Chem. Phys. Lett.~{\bf 73}, 393 (1980).

\bibitem{ccdiis}
G.E.~Scuseria, T.J.~Lee, H.F.~Schaefer III,
Chem. Phys. Lett.~{\bf 130}, 236 (1986).

\bibitem{dipea1}
M. Nooijen and R.J. Bartlett, J. Chem. Phys. {\bf 106}, 6441, 6812 (1997).

\bibitem{dipea2}
M. Wladyslawski and M. Nooijen, in {\em Low-Lying
Potential Energy Surfaces}, ACS Symposium Series, Vol. 828,
edited by M.R. Hoffmann and K.G. Dyall
(American Chemical Society, Washington, D.C., 2002), p. 65.

\bibitem{dipea3}
M. Nooijen, Int. J. Mol. Sci. {\bf 3}, 656 (2002).

\bibitem{lrcc0}
H.~Monkhorst, Int. J. Quantum Chem. Symp.~{\bf 11}, 421 (1977).

\bibitem{lrcc1}
K.~Emrich, Nucl. Phys. A {\bf 351}, 379 (1981).

\bibitem{lrcc2}
E.~Daalgard and H.~Monkhorst, Phys. Rev. A {\bf 28},
1217 (1983).

\bibitem{lrcc3}
M.~Takahashi and J.~Paldus, J. Chem. Phys.~{\bf 85}, 1486 (1986).

\bibitem{lrcc4}
H.~Koch and P.~J{\o}rgensen, J. Chem. Phys.~{\bf 93}, 3333, 3345 (1990).

\bibitem{sacci1}
H. Nakatsuji and K. Hirao, Chem. Phys. Lett.
{\bf 47}, 569 (1977).

\bibitem{sacci2}
H. Nakatsuji and K. Hirao, J. Chem. Phys. {\bf 68}, 2053, 4279 (1978).

\bibitem{sacci3}
H. Nakatsuji, Chem. Phys. Lett. {\bf 59}, 362 (1978).

\bibitem{sacci4}
H. Nakatsuji, Chem. Phys. Lett. {\bf 67}, 329, 334 (1979).

\bibitem{sacci5}
H. Nakatsuji, in {\em Computational Chemistry: Reviews of Current
Trends}, edited by J. Leszczy{\' n}ski (World Scientific, Singapore, 1997),
Vol. 2, p. 62.

\bibitem{sacciopen}
H. Nakatsuji and M. Ehara, J. Chem. Phys. {\bf 98}, 7179 (1993).

\bibitem{hirataeaip}
S. Hirata, M. Nooijen, and R.J. Bartlett, Chem. Phys. Lett.
{\bf 328}, 459 (2000).

\bibitem{vucc1}
I. Lindgren and D. Mukherjee, Phys. Rep. {\bf 151}, 93
(1987).

\bibitem{vucc2}
D. Mukherjee and S. Pal, Adv. Quantum Chem. {\bf 20},
291 (1989).

\bibitem{succ}
B. Jeziorski and H.J. Monkhorst, Phys. Rev. A {\bf 24},
1668 (1981).

\bibitem{pal}
K.R. Shamasundar and S. Pal, J. Chem. Phys. {\bf 114}, 1981 (2001);
{\bf 115}, 1979 (2001) (Erratum).

\bibitem{succ1}
K. Kowalski and P. Piecuch,
Phys. Rev. A {\bf 61}, 052506 (2000).

\bibitem{succ2}
P. Piecuch and J.I. Landman,
Parallel Comp. {\bf 26}, 913 (2000).

\bibitem{succ3}
K. Kowalski and P. Piecuch,
Chem. Phys. Lett. {\bf 334}, 89 (2001).

\bibitem{succ4}
K. Kowalski and P. Piecuch,
J. Molec. Struct.: THEOCHEM {\bf 547}, 191 (2001).

\bibitem{succ5}
P. Piecuch and K. Kowalski,
Int. J. Mol. Sci. {\bf 3}, 676 (2002).

\bibitem{succ6}
K. Kowalski and P. Piecuch,
Mol. Phys. {\bf 102}, 2425 (2004).

\bibitem{xlin1}
X. Li and J. Paldus, J. Chem. Phys. {\bf 119}, 5320 (2003).

\bibitem{xlin2}
X. Li and J. Paldus, J. Chem. Phys. {\bf 119}, 5334 (2003).

\bibitem{xlin3}
X. Li and J. Paldus, J. Chem. Phys. {\bf 119}, 5346 (2003).

\bibitem{xlin3a}
X. Li and J. Paldus, J. Chem. Phys. {\bf 120}, 5890 (2004).

\bibitem{fsccsdt1}
M. Musia{\l} and R.J. Bartlett,
J. Chem. Phys. {\bf 121}, 1670 (2004).

\bibitem{fsccsdt2}
M. Musia{\l}, L. Meissner, S.A. Kucharski, and R.J. Bartlett,
J. Chem. Phys. {\bf 122}, 224110 (2005).

\bibitem{succA}
J. Paldus, L. Pylypow, and B. Jeziorski,
in {\em Many-Body Methods in Quantum Chemistry},
Vol. 52, Lecture Notes in Chemistry, edited by U. Kaldor
(Springer, Berlin, 1989), p. 151.

\bibitem{succB}
P. Piecuch and J. Paldus,
Theor. Chim. Acta {\bf 83}, 69 (1992).

\bibitem{succC}
J. Paldus, P. Piecuch, B. Jeziorski, and L. Pylypow,
in {\it Recent Progress in Many-Body Theories}, Vol. 3,
edited by T. L. Ainsworthy, C. E. Campbell, B. E. Clements, and E. Krotschek
(Plenum, New York, 1992), Vol. 3, p. 287.

\bibitem{succD}
J. Paldus, P. Piecuch, L. Pylypow,
and B. Jeziorski, Phys. Rev. A {\bf 47}, 2738 (1993).

\bibitem{succE}
Piecuch and J. Paldus, Phys. Rev. A {\bf 49}, 3479 (1994).

\bibitem{succF}
P. Piecuch and J. Paldus, J. Chem. Phys. {\bf 101},
5875 (1994).

\bibitem{vuintruder1}
K. Jankowski, J. Paldus, I. Grabowski, and K. Kowalski,
J. Chem. Phys. {\bf 97}, 7600 (1992); {\bf 101}, 1759 (1994) [Erratum].

\bibitem{vuintruder2}
K. Jankowski, J. Paldus, I. Grabowski, and K. Kowalski, J. Chem. Phys.
{\bf 101}, 3085 (1994).

\bibitem{fujii2004}
S.~Fujii, R.~Okamoto, and K.~Suzuki, Phys.~Rev.~C {\bf 69}, 
054301(2004).

\bibitem{so84}
K.\ Suzuki, Prog.\ Theor.\ Phys.\ {\bf 68}, 1627 (1982); K.\ Suzuki and 
R.\ Okamoto, Prog.\ Theor.\ Phys.\ {\bf 75}, 1388 (1986); 
{\it ibid.} {\bf 76}, 127 (1986).

\bibitem{horoi:unpub}
M. Horoi, M. W{\l}och, M. Hjorth-Jensen, P. Piecuch, D.J. Dean, and B.A. Brown,
unpublished.

\bibitem{hirao}
K. Hirao and H. Nakatsuji, J. Comput. Phys. {\bf 45}, 246 (1982).

\bibitem{dav}
E.R. Davidson, J. Comput. Phys. {\bf 17}, 87 (1975).

\bibitem{audi2003}
G.~Audi, A.~H.~Wapstra, and C.~Thibault, Nucl.~Phys.~A {\bf 729}, 337 (2003).

\bibitem{navratil1999}
P.~Navr{\'a}til and B.R.~Barrett, Phys. Rev. C {\bf 59}, 1906 (1999);
P.~Navr{\'a}til, G.P.~Kamuntavicius, and B.R.~Barrett, Phys.~Rev.~C {\bf 61}, 044001 (2000).

\bibitem{ellis2005}
P.J.~Ellis, T.~Engeland, M.~Hjorth-Jensen, M.P.~Kartamyshev, and E.~Osnes,
Phys.~Rev.~C {\bf 71}, 034301 (2005).

\bibitem{epelbaum2003}
E.~Epelbaum, A.~Nogga, W.~Gl\"ockle, H.~Kamada, U.-G.~Meissner, and H.~Wita{\l}a,
Phys.~Rev.~C {\bf 66}, 064001 (2002).

\bibitem{fire}
R.B.~Firestone, V.S.~Shirley, C.M.~Baglin, S.Y.~Frank Chu,
and J.~Zipkin, Table of Isotopes, 8th ed.~(Wiley Interscience, New York, 1996).

\bibitem{comment}
Note however that within 
many-body perturbation theory,
this result strongly depends on the oscillator 
energy. With an unperturbed harmonic oscillator basis
the excitation energy corresponding to 
the $0p_{3/2}$ state varies from $2.2$ MeV
for $\hbar\Omega=10$ MeV to $10.1$ MeV for $\hbar\Omega=20$ MeV. 
These findings
agree with the similar findings of Fujii {\em et al.} 
\protect\cite{fujii2004}, demonstrating a clear
weakness of many-body perturbation theory. Coupled-cluster 
methods, in which selected
classes of diagrams are summed to infinite order, 
are much more robust in this regard.

\bibitem{andres1}
A.P.~Zuker, B.~Buck, and J.B.~McGrory , Phys.~Rev.~Lett.~{\bf 21}, 39 (1968).

\bibitem{pe70}
P.J.~Ellis and T.~Engeland, Nucl.~Phys.~A {\bf 144}, 161 (1970).

\bibitem{pe71}
T.~Engeland and P.J.~Ellis, Nucl.~Phys.~A {\bf 181}, 368 (1972).

\bibitem{hj90}
W.~Haxton and C.~Johnson, Phys.~Rev.~Lett.~ {\bf 65}, 1325 (1990).

\bibitem{wb92}
E.K.~Warburton and B.A.~Brown, Phys.~Rev.~C {\bf 46}, 923 (1992).

\bibitem{wbm92}
E.K.~Warburton, B.A.~Brown, and D.J.~Millener, Phys.~Lett.~B 
{\bf 293}, 7 (1992).

\bibitem{andres2}
A.P.~Zuker, Phys.~Rev.~Lett.~ {\bf 90 }, 042502 (2003).

\bibitem{andres3}
A.P.~Zuker, in {\em Key Topics in Nuclear Structure}, edited by A.~Covello,
(World Scientific, Singapore, 2005) p.~135.

\bibitem{steve2001}
S.C.~Pieper, V.R.~Pandharipande, R.B.~Wiringa, and J.~Carlson,
Phys.~Rev.~C {\bf 64}, 014001 (2001). 

\bibitem{gari1996}
J.A.~Eden and M.F.~Gari,
Phys.~Rev.~C {\bf 53}, 1510 (1996). 

\bibitem{witek}
N.~Michel, W. Nazarewicz, and M.~P{\l}oszajczak, 
Phys.~Rev.~C {\bf 70}, 064313 (2004). 

\bibitem{roberto}
R.~Id Betan, R.~Liotta, N.~Sandulescu, and T.~Vertse,
Phys.~Rev.~C {\bf 67}, 014322 (2003). 

\bibitem{gaute}
G.~Hagen, M.~Hjorth-Jensen, and J.S.~Vaagen,
Phys.~Rev.~C {\bf 71}, 044314 (2005). 

\end{thebibliography}
\end{document}